\documentclass[twocolumn,tighten]{aastex631}
\usepackage{longtable}

\hypersetup{linkcolor=violet,citecolor=blue,filecolor=magenta,urlcolor=cyan}

\newcommand{\Lx}{$L_{\rm X}$}
\newcommand{\Fx}{$F_{\rm X}$}
\newcommand{\lum}{erg s$^{-1}$}
\newcommand{\flux}{erg cm$^{-2}$ s$^{-1}$}
\newcommand{\xmm}{{\em XMM-Newton}}
\newcommand{\chandra}{{\em Chandra}}

\newcommand{\Msun}{$M_{\odot}$}
\newcommand{\Rsun}{$R_{\odot}$}

\newcommand{\Lsun}{$L_{\odot}$}
\newcommand{\Lbol}{$L_{\rm bol}$}

\newcommand{\vsini}{$v \sin i$}

\accepted{June 9, 2026}

\submitjournal{ApJ}

\begin{document}

\title{X-ray Emission and Stellar Ages of Sun-Like Stars}

\correspondingauthor{Breanna A. Binder}
\email{babinder@cpp.edu}

\author[0000-0002-4955-0471]{Breanna A. Binder}
\affiliation{Department of Physics and Astronomy, California State Polytechnic University, Pomona, CA, USA}

\author[0000-0002-2949-2163]{Edward W. Schwieterman}
\affiliation{Department of Earth and Planetary Sciences, University of California, Riverside, CA, USA}
\affiliation{Blue Marble Space Institute of Science, Seattle, WA, USA}

\author[0000-0003-0944-2334]{Alison Farrish}
\affiliation{NASA Goddard Space Flight Center, Greenbelt, MD 20771, USA}
\affiliation{George Mason University, Fairfax, VA 22030, USA}

\author[0000-0002-1046-025X]{Sarah Peacock}
\affiliation{NASA Goddard Space Flight Center, Greenbelt, MD 20771, USA}
\affiliation{University of Maryland, Baltimore County, Baltimore, MD, 21250, USA}

\author[0000-0002-0569-1643]{Margaret C. Turnbull}
\affiliation{SETI Institute, Carl Sagan Center for the Study of Life in the Universe, Off-Site: Madison, WI 53713, USA}

\author[0000-0002-7084-0529]{Stephen R. Kane}
\affiliation{Department of Earth and Planetary Sciences, University of California, Riverside, CA, USA}

\author[0000-0001-6398-8755]{Katherine Garcia-Sage}
\affiliation{Heliophysics Division, NASA Goddard Space Flight Center, Greenbelt, MD, USA}

\begin{abstract}
We present an analysis of \emph{XMM-Newton} and \emph{Chandra} observations of 85 nearby main-sequence FGK stars with age estimates ranging from 0.2-12 Gyr. We measure quiescent 0.3-10 keV luminosities, variability metrics, and multi-temperature thermal plasma spectral parameters. Quiescent spectra are typically described by three characteristic plasma components ($kT\approx0.1$, 0.4, 0.8 keV); the fraction of flux from $T\ge7$ MK rises with X-ray surface flux, reaching $\sim$50\% for $F_X\gtrsim10^6$ erg cm$^{-2}$ s$^{-1}$. We derive relations between emission measure-weighted coronal temperature and both $L_X$ and \Fx, enabling temperature-informed count-rate conversions for faint sources. We quantify how bandpass conversions (ROSAT 0.1-2.4 keV vs. \emph{XMM-Newton} 0.3-10 keV) depend on temperature, and show that inferred ROSAT-band $L_X$ broadly follows the canonical $t^{-1.5}$ decay, while the harder band exhibits increased scatter at $>$4 Gyr. Several stars show excess activity suggestive of age errors, inclination effects, or unresolved companions. Some of these ``outlier'' stars are potential direct imaging targets for the {\em Habitable Worlds Observatory}, and detailed characterization of these stars is needed to inform their likely influence on the atmospheric evolution of orbiting planets.
\end{abstract}

\section{Introduction}\label{sec:intro}
The ROSAT All-Sky Survey \citep[RASS;][]{Voges93,Voges+99}, along with surveys and targeted observations with other early X-ray telescopes such as the Einstein Observatory \citep{Vaiana+81}, revealed that nearly all nearby Sun-like main sequence stars (of spectral types F, G, and K) emit detectable X-ray radiation \citep{Schmitt97}, which originates from a hot ($\sim$few MK) stellar corona. Typical 0.1-2.4 keV luminosities (hereafter referred to as the RASS bandpass) for these stars span a broad range, $10^{26-30}$ \lum, with young and/or rapidly rotating stars producing more luminous X-ray emission and older, slower rotating stars showing diminished X-ray emission. Empirically, the X-ray luminosity is observed to decay with age \citep{Gudel+97_age} according to
\vspace{-0.1cm}
\begin{equation}\label{eq:Gudel04}
    L_X = (3\pm1)\times10^{28} t^{-1.5\pm0.3} \text{ erg s}^{-1},
\end{equation}

\noindent where $t$ is the age of the star in Gyr \citep[see also][]{Telleschi+05,Claire+12}. This decay in \Lx\ is believed to be the direct result of rotation rate spindown \citep{Skumanich72}, and therefore magnetic activity \citep{Shoda+20}, which is in turn tied to the stellar mass \citep{Gudel04,Matt+15}. Concurrent shifts in the X-ray spectral shape towards cooler plasma temperatures (with a rapid decrease in particular of the hot plasma component) are also expected in older stars \citep{Gudel04}, which could be the result of less efficient magnetic heating of the corona with age \citep{Gudel+97_age,Gudel97_flares}. RASS additionally helped to establish baseline X-ray-to-bolometric luminosity ratios (\Lx/\Lbol) for stars of different spectral types; generally, the lowest levels of X-ray activity in FGK stars are observed at \Lx/\Lbol$\sim10^{-8}$ \citep[see][and references therein]{Gudel04,Testa+15,Ayres25} in old and slowly rotating stars, while \Lx/\Lbol\ saturates at $\approx10^{-3}$ for fast rotating, young ($<$a few Myr) stars \citep{Wright+11}.

The dominant plasma in the Solar corona reaches temperatures of a $\sim$few MK, with active regions and flaring events producing coronal regions reaching $\sim$10-20 MK. The higher X-ray sensitivity and expanded energy ranges (up to $\sim$10 keV) of \xmm\ and \chandra\ enabled detailed study of the coronal plasmas of other nearby FGK stars, which also exhibited ``cool'' ($\sim$1-3 MK) plasma components that dominated the X-ray emission of older stars and ``hot'' ($\sim$10-20 MK) plasma components detected in younger stars and flaring events \citep{Soderblom10,Gudel04,Testa+15}. Statistical analyses of X-ray variability in young stars showed that a continuum of frequent flaring events could be responsible for steady heating of the corona \citep{Stelzer+07}. Empirically, the emission-measure weighted coronal temperature $T$ is related to the surface flux \Fx\ as $T\propto0.11$\Fx$^{0.26}$ \citep{Johnstone+15,Wood+18}, where $T$ is measured in MK and \Fx\ in \flux. The concurrent decline in both \Fx\ and the coronal temperature with age further implicates the important role of magnetic activity in the generation of X-rays from these stars. 

Understanding the age-related X-ray characteristics of FGK stars is crucial for predicting the atmospheric properties of orbiting planets and interpreting data from future exoplanet direct-imaging missions. The XUV radiation from the host star heats the upper atmosphere of orbiting planets, leading to the removal of neutral gas through thermal escape processes \citep{Tian2008, Erkaev2013,Johnstone2021b}. Some models indicate that an Earth-like N$_{2}$-dominated atmosphere can be lost to space within a few million years at XUV fluxes only five to six times higher than those received by modern Earth, due to thermospheric heating and subsequent atmospheric erosion, depending on atmospheric composition and the abundance of coolant molecules such as CO$_{2}$ \citep{Tian2008,Johnstone2021b}. XUV radiation also drives non-thermal escape by ionizing atmospheric species, enhancing ionospheric densities, and facilitating ion pickup by the stellar wind (including CME events) and polar outflows \citep{Lammer2007, Airapetian2017,GarciaSage+17,Scherf2021}. To maintain an atmosphere, a terrestrial planet must replenish these losses through volcanic outgassing, exogenous delivery, or, in the case of inhabited planets, biological activity. Understanding the intersection of scatter in expected XUV activity of stars with age, along with the corresponding scatter in the expected geothermal and volcanic outgassing lifetimes of terrestrial planets, would therefore improve our understanding of the likely atmospheric evolution histories of terrestrial exoplanets. It also allows us to develop testable scientific hypotheses in advance of statistical surveys through volume-limited direct-imaging missions like the Habitable Worlds Observatory. An improved calibration of stellar ages would likewise contribute to this goal by providing planetary ages, potentially improving our ability to test proposed statistical biosignatures. One example is the age-oxygen relationship \citep{Bixel2020, Blunt2025}, which is the hypothesis that older habitable planets should be more likely to show O$_{2}$ and/or O$_{3}$ spectral signatures from photosynthesis due to the time necessary to evolve oxygenic photosynthesis and oxygenate the planetary environment \citep{Lyons2014,schwieterman+2018}. These factors together motivate collecting an expanded sample of XUV activity with age for Sun-like stars to validate or refine existing relationships and identify age errors for future direct-imaging targets. 

Recent models by \citet{Johnstone2021} suggest that all $\sim$4-5 Gyr FGK stars should have coronal temperatures of $\sim$2.5 MK regardless of the initial spin rate. However, the relative faintness and lack of deep, uniform X-ray data sets of stars older than a few Gyr make it difficult to constrain models of X-ray emission for older Sun-like stars. We address this by extending their analysis to a larger sample of 85 nearby FGK field stars with age estimates and targeted \xmm\ and/or \chandra\ observations. Our sample spans ages 0.2-12 Gyr, of with 37 stars ($\sim$44\% of the sample) are $\geq$4 Gyr. In Section~\ref{section:sample} we discuss the sample selection criteria, compile available stellar parameters, and quantify the X-ray coverage of stars in the sample by \xmm\ and \chandra. In Section~\ref{sec:analysis} we provide a brief description of our analysis methodology and present the results of our X-ray spectral analysis.  In Section~\ref{sec:age_dependence}, we discuss the dependence of the observed X-ray properties (e.g., \Lx, surface flux) on stellar age, and we finally present conclusions of our work in Section~\ref{sec:conclusion}.

\section{Sample Selection}\label{section:sample}

Stellar ages are notoriously difficult to measure, particularly for low-mass main sequence stars \citep{Soderblom10}. While the ages of stellar clusters can be determined by measuring the main sequence turn-off observed on a Hertzsprung-Russell (H-R) diagram, field stars lack coeval stellar populations that might enable age dating via H-R diagram analysis. Instead, the ages of Sun-like stars have historically been inferred through measurements of the Ca~II H and K absorption lines, which probe chromospheric activity generated via the stellar magnetic dynamo \citep{Soderblom10}. The Ca II infrared triplet at 8498.062 (T1), 8542.144 (T2), and 8662.170 (T3) additionally probe stellar activity at a broader range of atmospheric layers \citep{Chmielewski00,Busa+07}. The H$\alpha$ line, which is less sensitive to stellar magnetic cycles and related magnetic activity (e.g., sunspots), has emerged a useful diagnostic of stellar age when used in conjunction with Ca II H and K lines \citep{Souza+24}; it also has the benefit of being almost entirely insensitive to stellar metallicity \citep{Fuhrmann+93}. Aside from individual spectral lines, stellar spectra in their entirety can be compared to grids of theoretical stellar models computed from stellar evolution codes to constrain ages. Isochrone modeling and X-ray luminosity, too, are known to track ages, and significant effort has been put into compiling multiple indicators into a single age determination for Sun-like stars. 

We caution that activity-based age indicators are not fully independent of the X-ray properties; ages inferred from chromospheric line emission, rotation, or \Lx\ may encode some of the same magnetic-activity evolution that we seek to characterize in this work. Thus, using such ages to calibrate X-ray evolution can introduce a degree of circularity, particularly for stars whose adopted ages rely heavily on activity diagnostics. To minimize this effect, we compile ages from multiple literature methods where available and adopt the median age estimate. We interpret discrepancies between X-ray activity level and adopted age (e.g., see Section~\ref{sec:outliers}) as potential evidence for age uncertainty, inclination effects, unresolved companions, or residual unresolved X-ray variability.

To define our sample of target stars, we first begin by searching for F, G, and K-type stars with age estimates from one of four sources: (1) \citet{Mamajek+08}, which analyzed a combination Ca II H and K lines, X-ray luminosities from ROSAT, and gyrochronology studies \citep[using rotation periods and $B-V$ colors; e.g.,][]{Barnes07} to determine or refine age estimates of 108 solar-type field dwarfs within 16 pc, with particular attention paid to stars with ages $<$2 Gyr{\footnote{Gyrochronology ages are uncertain for $\lesssim$1 Gyr field stars because rotational convergence is incomplete; stars of similar mass can retain a broad range of rotation periods until several hundred Myr to $\sim$1 Gyr \citep{Tu+15,Johnstone2021}.}
;
(2) \citet{Takeda+07}, which derived theoretical spectral models for 1074 nearby stars from the Spectroscopic Properties of Cool Stars (SPOCS) catalog using the Yale Rotational Evolution Code (YREC); (3) \citet{Lorenzo+16}) and \citet{Lorenzo+18}, which analyzed the Ca II H and K and infrared triplet activity indicators to estimate ages for 113 FGK stars, with an emphasis on expanding age-activity relations to stars older than $\sim$2 Gyr; and (4) \citet{Souza+24}, which calibrated H$\alpha$ line fluxes with Ca II H+K activity indicators to provide age estimates for 511 solar-type stars \citep[inclusive of the stars analyzed in][]{Lorenzo+16,Lorenzo+18}.

\small
\setlength{\tabcolsep}{4pt}
\begin{longtable*}{llcccccccccccc}
    \caption{Nearby FGK Stars Not in \citet{Binder+24} with Targeted X-ray Observations}\label{tab:stellar_sample} \\
    \hline \hline
    HD      & Alt           & Distance  & Spectral  & Mass      & Radius    & \Lbol     & $T_{\rm eff}$ & $v$sin$i$  & \multicolumn{2}{c}{\xmm} && \multicolumn{2}{c}{\chandra} \\ \cline{10-11} \cline{13-14}
    \#  & Name          & (pc)      & Type      & (\Msun)   & (\Rsun)   & (\Lsun)   & (K) & (km s$^{-1}$) & \# Obs & Time (ks) && \# Obs & Time (ks) \\
    \hline
    \endfirsthead

    \multicolumn{14}{c}{{\bfseries \tablename\ \thetable{} -- continued from previous page}} \\ \hline
    HD      & Alt           & Distance  & Spectral  & Mass      & Radius    & \Lbol     & $T_{\rm eff}$ & $v$sin$i$  & \multicolumn{2}{c}{\xmm} && \multicolumn{2}{c}{\chandra} \\ \cline{10-11} \cline{13-14}
    \#  & Name          & (pc)      & Type      & (\Msun)   & (\Rsun)   & (\Lsun)   & (K) & (km s$^{-1}$) & \# Obs & Time (ks) && \# Obs & Time (ks) \\
    \hline
    \endhead

    \hline \multicolumn{14}{r}{{\bf Continued on next page}} \\
    \endfoot

    \hline \hline

    \multicolumn{14}{p{\dimexpr\linewidth-2\tabcolsep\relax}}{\footnotesize
    $^*$452 Vul was observed multiple times by \xmm, but these observations suffer from significant blending with nearby sources (which are easily resolved by \chandra).}\\   
    \endlastfoot
    1461    & GJ 16.1       & 23.45 & G3V   & 1.03  & 1.15  & 1.24  & 5695  & 1.6   & 1  & 36.4    && 0 & \nodata  \\
    1835    & 9 Cet         & 21.30 & G5V   & 1.04  & 1.03  & 1.09  & 5794  & 7.0   & 1  & 6.9     && 0  & \nodata  \\
    3651    & 54 Psc        & 11.13 & K0V   & 0.89  & 0.87  & 0.53  & 5206  & 1.1   & 1  & 9.0 && 0 & \nodata  \\
    3765    &               & 17.93 & K2.5V & 0.85  & 0.83  & 0.38  & 5050  & 2.0   & 2 & 49.4 && 0 & \nodata \\
    8673    &               & 37.86 & F5V   & 1.28  & 1.50  & 3.39  & 6389  & 26.9  & 1 & 52.2 && 0 & \nodata \\
    11505   & HIP 8798      & 38.97 & G5V   & 1.03  & 1.20  & 1.42  & 5757  & 1.5   & 1  & 27.1 && 0 & \nodata  \\
    25680   & 39 Tau        & 16.89 & G1V   & 1.06  & 1.00  & 1.05  & 5848  & 3.8   & 1  & 3.7     && 0  & \nodata  \\
    25874   & HIP 18844     & 25.95 & G2V   & 1.02  & 1.11  & 1.19  & 5735  & 0.7   & 1  & 8.7 && 0 & \nodata  \\
    27442   & $\epsilon$ Ret & 18.27 & K2III & 1.48 & 3.54  & 7.09  & 4851  & 2.1   & 2 & 33.5 && 0 & \nodata \\
    28099   & HIP 20741     & 45.88 & G0V   & 1.04  & 1.01  & 1.04  & 5804  & 4.0   & 1 & 7.5 && 0 & \nodata  \\
    28946   &               & 27.73 & G9V   & 0.92  & 0.83  & 0.49  & 5338  & 2.1   & 4 & 219.7 && 0 & \nodata \\
    33636   &               & 28.69 & G0V   & 1.02  & 1.02  & 1.15  & 5904  & 3.1   & 1 & 6.8 && 0 & \nodata \\
    38529   & 138 G. Ori    & 42.35 & G4V   & 0.98  & 2.84  & 6.82  & 5541  & 3.2   & 1 & 11.3 && 0 & \nodata \\    
    61421   & Procyon A     & 3.51  & F5IV/V & 1.49 & 1.98  & 7.11  & 6584  & 5.7   & 5 & 203.5 && 0 & \nodata  \\
    63433   &               & 22.40 & G5V   & 0.99  & 0.91  & 0.76  & 5688  & 7.3   & 1 & 5.1 && 0 & \nodata \\
    75289   &               & 29.12 & F9V   & 1.12  & 1.30  & 2.04  & 6044  & 4.1   & 2 & 20.3 && 0 & \nodata \\ 
    79969   & LHS 2125      & 17.53 & K3V   & 0.83  & 1.08  & 0.47  & 4825  & 6.0   & 1 & 14.9 && 0 & \nodata \\
    82443   & GJ 354.1      & 18.07 & K1V   & 0.91  & 0.83  & 0.47  & 5287  & 10.4  & 1 & 10.1 && 7 & 104.5  \\
    82558   &               & 18.28 & K1V   & 0.84  & 0.69  & 0.27  & 5026  & 26.6  & 2 & 73.9 && 0 & \nodata \\
    99491   & 83 Leo A      & 18.20 & G8III/IV & 0.95 & 0.99 & 0.77 & 5438  & 1.4   & 1  & 21.4    && 0  & \nodata \\
    102117  &               & 39.58 & G6V   & 1.01  & 1.24  & 1.44  & 5668  & 0.9   & 0 & \nodata && 1 & 7.0 \\
    108147  &               & 38.92 & F8V   & 1.22  & 1.21  & 2.01  & 6256  & 6.1   & 4 & 48.6 && 0 & \nodata \\
    108309  &               & 26.67 & G2V   & 1.05  & 1.39  & 2.02  & 5778  & 1.2   & 4 & 126.9 && 0 & \nodata  \\
    114386  &               & 27.92 & K3V   & 0.78  & 0.77  & 0.29  & 4827  & 0.6   & 1 & 1.7 && 0 & \nodata \\
    114783  &               & 21.06 & K1V   & 0.90  & 0.84  & 0.42  & 5127 & 2.0    & 1 & 8.5 && 0 & \nodata \\
    120066  & HR 5183       & 31.46 & G0V   & 1.07  & 1.56  & 2.48  & 5794  & 3.5   & 0 & \nodata && 1 & 23.0 \\
    120136  & $\tau$ Boo    & 15.65 & F7V   & 1.40  & 1.44  & 3.20  & 6461  & 15.0  & 9 & 120.4 && 2 & 10.1  \\
    120690  & GJ 530        & 18.55 & G5V   & 1.01  & 0.94  & 0.82  & 5663  & 2.7   & 1 & 9.5 && 0 & \nodata \\
    129333  & EK Dra        & 34.40 & G5V   & 1.04  & 0.97  & 0.92  & 5744  & 16.8  & 1 & 50.0 && 0 & \nodata \\
    130307  & GJ 3867       & 19.67 & K2.5V & 0.78  & 0.75  & 0.31  & 5024  & 2.0   & 2 & 6.2 && 0 & \nodata \\
    135599  & HIP 74702     & 15.58 & K0V   & 0.82  & 0.80  & 0.43  & 5221  & 3.6   & 1 & 6.6 && 0 & \nodata \\
    145417  &               & 13.61	& K3V	& 0.83	& 0.58	& 0.19	& 4993	& 4.0   & 0 & \nodata && 1 & 35.2 \\
    150248  & HIP 81746     & 27.77 & G3V   & 1.02  & 1.04  & 1.05  & 5715  & 0.8   & 2 & 43.2 && 0 & \nodata  \\
    157214  & 72 Her        & 14.54 & G0V   & 1.05  & 1.12  & 1.29  & 5817  & 1.7   & 0  & \nodata && 1  & 10.1 \\
    157347  & HIP 85042     & 19.46 & G3V   & 0.98  & 1.02  & 1.07  & 5714  & 0.9   & 1 & 33.8 && 0 & \nodata  \\
    162020  &               & 30.82	& K3V	& 0.77	& 0.74	& 0.26	& 4776	& 1.5   & 2 & 74.0 && 0 & \nodata \\
    165567  &               & 52.24	& F5V	& 1.31	& 1.94	& 5.48	& 6338	& 13.6  & 1 & 20.0 && 0 & \nodata \\
    170657  & GJ 716        & 13.21 & K2V   & 0.81  & 0.76  & 0.35  & 5103  & 1.5   & 1 & 27.8 && 0 & \nodata \\
    179949  & GJ 749        & 27.46 & F8.5V & 1.22  & 1.27  & 1.99  & 6176  & 6.5   & 15 & 282.0 && 5 & 152.2  \\
    187237  & HIP 97420     & 26.01 & G2IV/V & 1.05 & 1.00  & 1.10  & 5831  & 2.9   & 1 & 42.9 && 0 & \nodata   \\
    189733  & V452 Vul      & 19.76 & K2V   & 0.83  & 0.78  & 0.33  & 5012  & 3.5   & 0$^*$ & \nodata && 6 & 124.5 \\
    190771  & GJ 1249       & 19.01 & G2V   & 1.03  & 1.04  & 1.07  & 5767  & 4.3   & 1 & 25.7 && 0 & \nodata \\
    192020  &               & 24.46	& G8V	& 0.86	& 0.80	& 0.42  & 5181	& 1.0   & 1 & 31.3 && 0 & \nodata \\
    193664  & GL 778        & 17.57 & G0V   & 1.02  & 1.03  & 1.18  & 5886  & 1.3   & 1 & 1.0 && 0 & \nodata \\
    207740  &               & 49.25 & G5V   & 0.92  & 1.09  & 1.17  & 5746  & 10.5  & 1 & 1.8 && 0 & \nodata \\
    209779  & NT Aqr        & 36.45 & G1V   & 0.96  & 1.08  & 1.14  & 5739  & 10.2  & 1 & 4.3 && 0 & \nodata \\
    210918  & GJ 851.2      & 22.13 & G2V   & 0.96  & 1.19  & 1.42  & 5735  & 0.4   & 1 & 14.0 && 0 & \nodata  \\ 
    217107  & HIP 113421    & 20.06 & G8IV  & 1.06  & 1.17  & 1.25  & 5606  & 0.5   & 1 & 6.0 && 0 & \nodata 
    \label{tab:sample}
\end{longtable*}

We next searched the \xmm\footnote{See \url{https://nxsa.esac.esa.int/nxsa-web/\#home}} and \chandra\footnote{See \url{https://cda.harvard.edu/chaser/}} archives for observations of any star with an age estimate from one or more of the sources described above. We restricted \chandra\ observations to those obtained in imaging mode with ACIS-I or ACIS-S so that time-resolved spectroscopy could be performed (see Section~\ref{sec:analysis} for further details). We found 85 stars (excluding systems in close binaries that are unresolvable in the available X-ray observations) with at least one targeted X-ray observation; 37 of these stars were included in \citet{Binder+24}, which specifically examined the X-ray properties of stars of high interest for future exoplanet direct imaging missions. We therefore present an analysis of the X-ray data for the remaining 48 stars in this work, and use the already-available data products for the remaining stars from \citet{Binder+24}. We retrieved a total of 111 observations (87 from \xmm\ and 24 from \chandra) totaling $\sim$2.3 Msec in effective exposure time ($\sim$80\% of the total exposure time was obtained by \xmm). Table~\ref{tab:stellar_sample} presents a summary of the stars analyzed for this work. Distances were taken from {\em Gaia} DR3 \citep{GaiaDR3}. Spectral types were taken from \citet{Gray+06}. The spectral parameters of stellar mass, radius, bolometric luminosity, and effective temperature were compiled from the Revised TESS input catalog \citep{TESSinput}, ExoCat \citep{Turnbull15}\footnote{See \url{https://nexsci.caltech.edu/missions/EXEP/EXEPstarlist.html}}, and \citet{Takeda+07}, and \vsini\ values were adopted from \citet{Valenti+05}. 

Out of the 85 FGK stars in our sample, 33 stars ($\sim$39\% of the sample) had only one age estimate available from one of the four works referenced above. In these instances, we adopted the only age estimate available. For the remaining 52 stars ($\sim$61\% of the sample), we adopted the median age (which, for the majority of stars, did not significantly differ from the mean age) from the available age estimates and assume a $\sim$30\% age uncertainty \citep[similar to the typical uncertainty reported in both][]{Mamajek+08,Takeda+07}. Table~\ref{tab:stellar_ages} lists the reported ages for all the stars in our sample from \citet[][the ``composite'' age]{Mamajek+08}, \citet[][based on the YREC models]{Takeda+07}, the \citet[][Ca~II H+K]{Lorenzo+16,Lorenzo+18}, and the \citet[H$\alpha$ line fluxes]{Souza+24}, along with the age we adopted in this work. Figure~\ref{fig:age_comparison} compares the ages of stars derived via multiple methods, and Figure~\ref{fig:adopted_age} illustrates how the ages adopted in this work compare to the ages derived by the four different methods. Figure~\ref{fig:Age_dist} shows the distribution of stellar ages (by spectral type) for stars in our sample. 

\begin{deluxetable}{lcccccc}
    \tablecaption{Stellar Ages (in Gyr)}\label{tab:stellar_ages}
\tablehead{ 
          & Composite   & YREC  & H$\alpha$  & Ca~II H+K      & Adopted \\
    HD    & [1]         & [2]   & [3]        & [4]             & Age
 }
    \startdata 
    1461    & \nodata       & 7.1$^{+1.4}_{-1.6}$   & 3.2$^{+2.4}_{-1.4}$   & 5.0$^{+1.9}_{-1.4}$   & 5.0$\pm$1.5 \\
    1835    & \nodata       & $<$1.8                & 1.3$^{+0.9}_{-0.5}$   & 0.4$^{+0.2}_{-0.1}$   & 0.8$\pm$0.3 \\
    2151*   & \nodata       & 6.3$^{+0.3}_{-0.2}$   & 7.3$^{+5.4}_{-3.1}$   & 5.7$\pm$2.2           & 6.3$\pm$1.9 \\
    3651    & 7.7$\pm$0.5   & $>$11.8               & 6.2$^{+4.6}_{-2.6}$   & 6.0$^{+2.3}_{-1.7}$   & 6.2$\pm$1.9 \\
    3765    & \nodata       & 2.0$\pm$0.9           & \nodata               & \nodata               & 2.0$\pm$0.9 \\
    7570*   & 2.8$\pm$0.8   & 2.9$^{+1.2}_{-1.0}$   & 6.4$^{+4.7}_{-2.7}$   & 3.9$^{+1.5}_{-1.1}$   & 3.4$\pm$1.0 \\
    8673    & \nodata       & 2.5$\pm$0.2           & \nodata               & \nodata               & 2.5$\pm$0.2 \\
    9826*   & \nodata       & 3.1$\pm$0.2           & \nodata               & \nodata               & 3.1$\pm$0.2 \\
    10700*  & 5.8$\pm$1.7   & $>$12.1               & \nodata               & \nodata               & 5.8$\pm$1.7 \\
    \enddata
\tablecomments{[1] \citet{Mamajek+08}, [2] \citet{Takeda+07}, [3] \citet{Souza+24}, [4] \citet{Lorenzo+16,Lorenzo+18}; *X-ray properties discussed in \citet{Binder+24}. Only the first ten entries are shown; the full machine-readable table will be made available online from the journal.}
\end{deluxetable}

\begin{figure*}
    \centering
    \includegraphics[width=0.85\linewidth]{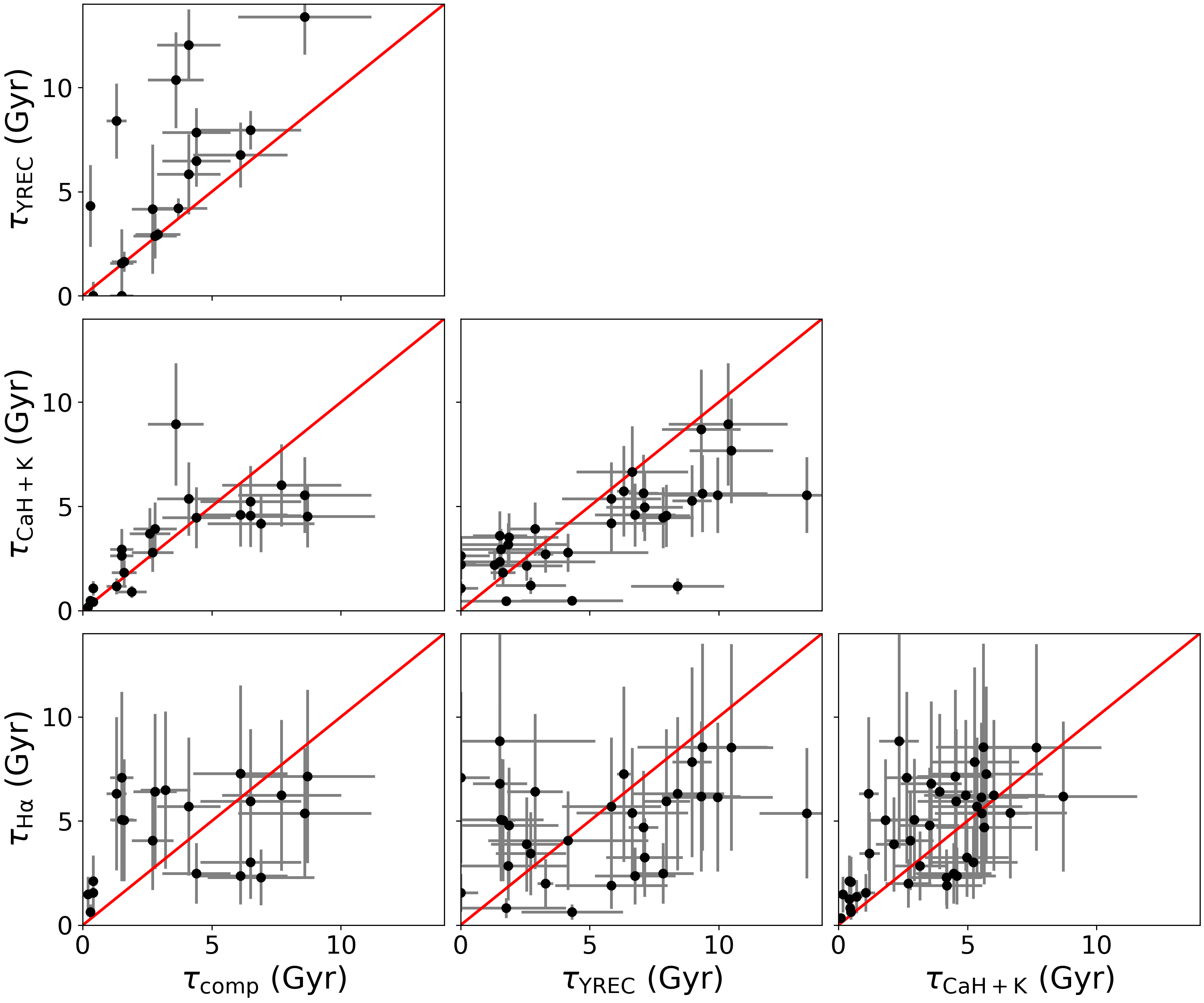}
    \caption{Comparison of stellar ages derived from different methods for stars with more than one age estimate. Composite ages ($\tau_{\rm comp}$) are taken from \citet{Mamajek+08}, SPOCS stars with stellar parameters (including ages) derived from YREC are taken from \citet[][$\tau_{\rm YREC}$]{Takeda+07}, Ca~II H+K age estimates are taken from \citet{Lorenzo+16,Lorenzo+18}, and age estimates from H$\alpha$ line fluxes are from \citet{Souza+24}. The red line shows a one-to-one correspondence.}
    \label{fig:age_comparison}
\end{figure*}

\begin{figure}
    \centering
    \includegraphics[width=0.85\linewidth]{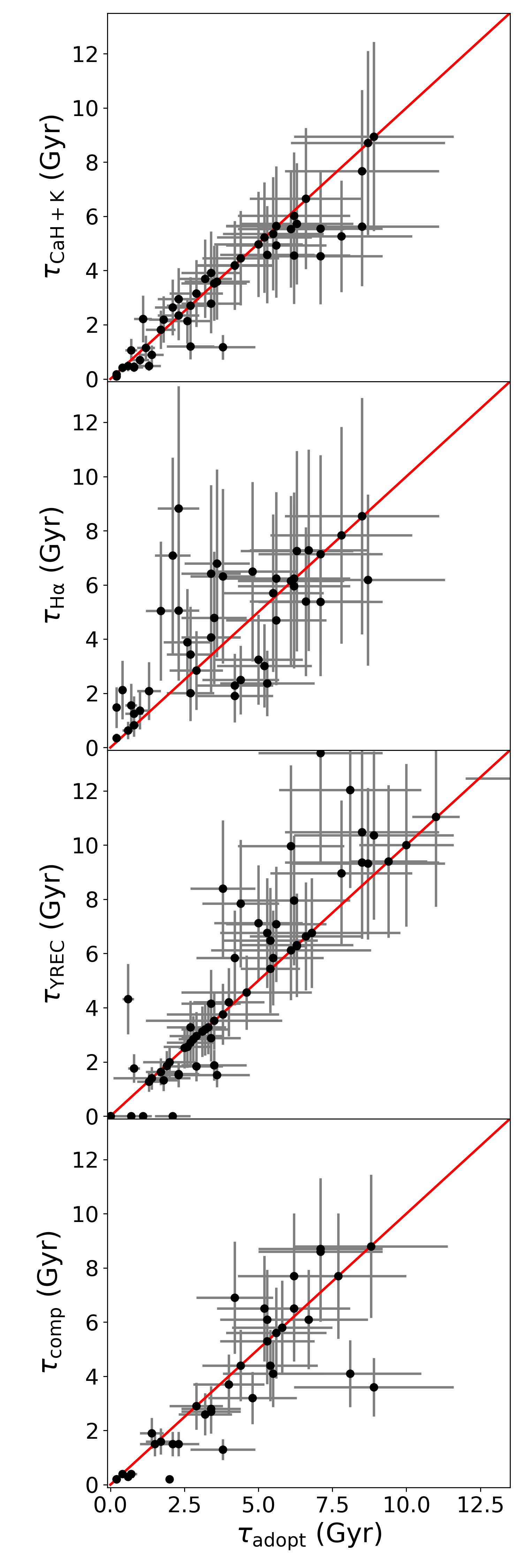}
    \caption{Our adopted age compared to the those derived from the four methods described in the text. The red line shows a one-to-one correspondence.}
    \label{fig:adopted_age}
\end{figure}

\begin{figure}
    \centering
    \includegraphics[width=1\linewidth]{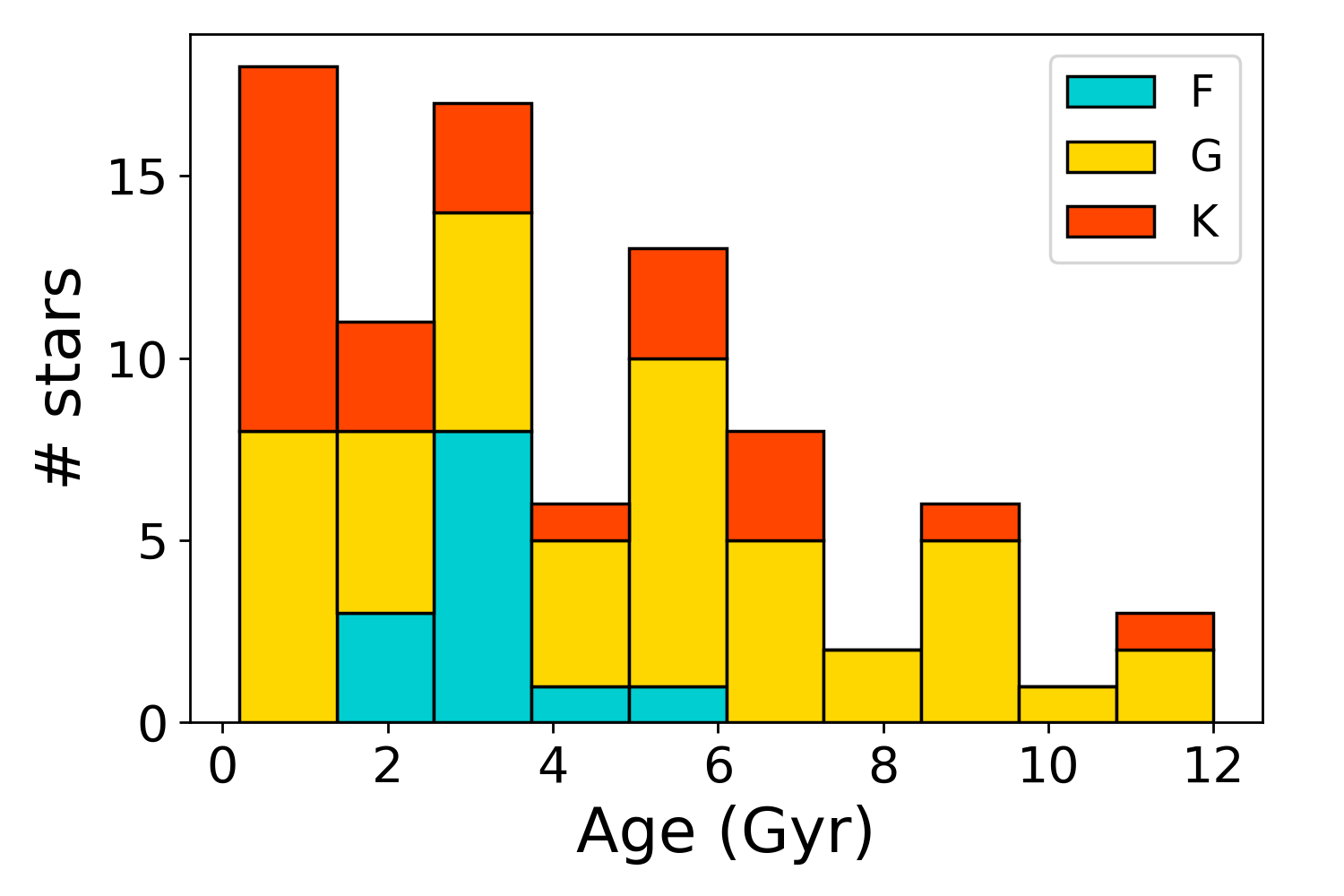}
    \caption{Distribution of ages for stars in our sample. The F star distribution is shown in cyan, the G star distribution is in yellow, and the K star distribution is in red-orange.}
    \label{fig:Age_dist}
\end{figure}

\vspace{-0.35cm}
\section{X-ray Data Analysis}\label{sec:analysis}
We follow the same general data reduction and analysis techniques described in \citet{Binder+24}. We provide a summary of the methodology here, and the reader is referred to \citet{Binder+24} for further details.

\xmm\ data were reprocessed from the raw \texttt{evt1} files using SAS 22.1\footnote{See \url{https://www.cosmos.esa.int/web/xmm-newton/sas}} and standard reduction procedures. The PN and MOS data were reprocessed with \texttt{epproc} and \texttt{emproc}, respectively, and the energy range was restricted to 0.2-15 keV. We extracted and inspected background light curves, then filtered the observations on good time intervals (i.e., we rejected periods of strong background flares) to create ``clean'' \texttt{evt2} files. These \texttt{evt2} files were then inspected to determine if an X-ray source was found coincident with our target star. If the target star was detected with $\gtrsim$500 net counts in the PN image, we extracted light curves (binned to 100 s). For stars with $\gtrsim$2000 net counts in the observation (or in specific time intervals within a single observation) we extracted spectra using \texttt{evselect}. Redistribution matrix files (RMFs) and ancillary response functions (ARFs) were generated using \texttt{rmfgen} and \texttt{arfgen}, respectively, and spectra were binned to contain $\geq$25 counts per bin to enable the application of $\chi^2$ statistics.

\chandra\ observations were similarly reprocessed from the raw \texttt{evt1} data with CIAO v4.17 \citep{Fruscione+06} and CALDB v4.12.0 using the \texttt{chandra\_repro} task and standard reduction procedures. The \chandra\ datasets utilized in this work are contained in the Chandra Data Collection~\dataset[DOI: 10.25574/cdc.598]{https://doi.org/10.25574/cdc.598}. Background light curves were again extracted and good time intervals were identified using the task \texttt{lc\_clean}. The reprocessed \texttt{evt2} data were then filtered on these good time intervals and the energy range was restricted to 0.5-7 keV. The CIAO tools \texttt{dmextract} and \texttt{specextract} were used to extract light curves (for sources detected with $\gtrsim$50 net counts) and spectra (for sources with $\gtrsim$500 net counts), respectively, in analogy with the \xmm\ observations described above.

\citet{Binder+24} utilized the Anderson-Darling statistic $A^2$ \citep{Feigelson+22} using \texttt{scipy}'s \texttt{anderson} routine and the reduced $\chi^2$ of the light curve compared to a constant count rate to assess whether count rate variability was detected in the binned light curves for each observation. We compute these metrics for each light curve that we are able to extract. We then visually inspect each light curve and identify time intervals containing strong flaring events (flagged as ``F''), monotonically rising (``R'') or diminishing (``D'') count rates, and quiescent (``Q'') intervals for each star. We also note intervals as elevated (“E”) when the count rate is $\sim$50\% above the lowest stable (quiescent) count rate level identified in the observation, but does not show the sharp rising or decaying morphology associated with a strong flare. The quiescent level is defined from intervals where the count rate is low and consistent with being constant within the uncertainties \citep{Binder+24}. In practice, the distinction between quiescent (``Q'') and elevated (``E'') intervals is limited by photon statistics; for short observations and/or faint sources, the count rate uncertainties are often sufficiently large such that modest elevations in count rate states cannot be robustly distinguished from the quiescent level. In such cases, any low-level elevated emission may be blended into the interval classified as ``quiescent.'' We therefore only assign the ``E'' flag when the source is sufficiently bright, or the observation sufficiently long, that two count rate regimes can be clearly distinguished. Unless otherwise noted, all X-ray luminosities (\Lx) and surface fluxes (\Fx) reported in this work refer to those measured during quiescent times when no strong count rate variability is observed.

\subsection{X-ray Spectroscopy}
For sufficiently bright stars ($\gtrsim$5000 net counts for \xmm\ observations and $\gtrsim$2000 net counts for \chandra\ observations), we extracted spectra in time intervals corresponding to the different variability flags described above. We used XSPEC \citep[v12.15.0;][]{Arnaud96} and standard $\chi^2$ statistics to model all the spectra using a one-, two-, or three-temperature thermal plasma (APEC) model subject to a minimal Galactic absorbing column (TBABS, with $n_H$ fixed to $10^{19}$ cm$^{-2}$). The plasma temperatures, $kT$ (in units of keV), and normalizations (which are related to the emission measure of the model component) were left as free parameters. The global APEC abundance parameter, $Z$, was fixed to the solar value ($Z=1$) unless a statistically acceptable fit could not be obtained with solar abundance and the spectrum had sufficient signal-to-noise to meaningfully constrain $Z$. For some spectra, allowing Z to vary produced poorly constrained or unphysical values, while the best-fit temperatures and normalizations remained insensitive to the assumed abundance. We therefore assumed $Z=1$ as the default, and allowed $Z$ to be a free parameter in cases where the spectrum had sufficient signal-to-noise to meaningfully constrain the $Z$, a statistically acceptable fit could not be obtained assuming Solar abundances, or the plasma temperatures and/or normalizations showed sensitivity to the value of $Z$. For stars with multiple observations and/or multiple time intervals representing the same general variability state, we computed the average of the the best-fit parameter values. All the best-fit spectral parameters for the sufficiently bright stars in our sample are summarized in Table~\ref{tab:avg_spectra}. Uncertainties correspond to the 90\% confidence interval.

\begin{deluxetable*}{cccccccccccccccccc}
\tabletypesize{\footnotesize}
\setlength{\tabcolsep}{2pt}
    \tablecaption{Average Best-Fit Spectral Parameters}\label{tab:avg_spectra}
\tablehead{
        & & \multicolumn{3}{c}{APEC \#1} && \multicolumn{3}{c}{APEC \#2} && \multicolumn{3}{c}{APEC \#3} && \\ \cline{3-5} \cline{7-9} \cline{11-13}
          & Var- & $kT$ & $\mathcal{N}$ & log$F_X$ [erg && $kT$  & $\mathcal{N}$ & log$F_X$ [erg  && $kT$ & $\mathcal{N}$ & log$F_X$ [erg & Abun- & average & log\Lx\  \\ 
    \colhead{HD} & Flag   & (keV) & ($\times10^{-4}$) & s$^{-1}$ cm$^{-2}$] &&  (keV) & ($\times10^{-4}$) & s$^{-1}$ cm$^{-2}]$ &&  (keV) & ($\times10^{-4}$) & s$^{-1}$ cm$^{-2}$] & dance\tablenotemark{{\bf \dag}} & $\chi^2_r$ & [\lum]
 }
    \startdata 
    1835 & Q & 0.12$^{+0.04}_{-0.02}$ & 2.6$^{+1.7}_{-1.2}$ & -12.79 && 0.50$^{+0.05}_{-0.10}$ & 3.9$^{+0.3}_{-1.3}$ & -12.69 && 1.00$^{+0.53}_{-0.60}$ & 0.6$^{+1.3}_{-0.3}$ & -11.93 & 1 & 1.24 & 28.92 \\
    8673 & Q & \nodata & \nodata & \nodata && 0.58$^{+0.03}_{-0.04}$ & 5.5$^{+0.3}_{-0.2}$ & -12.39 && \nodata & \nodata & \nodata & 0.13 & 1.10 & 28.85 \\
    28946 & Q & \nodata & \nodata & \nodata && 0.63$^{+0.07}_{-0.08}$ & 695.1$^{+143.3}_{-111.4}$ & -10.56 && 9.83$^{+1.41}_{-1.36}$ & 402.2$^{+56.2}_{-55.7}$ & -10.16 & $<$0.01 & 1.48 & 30.95 \\
    28946 & D & \nodata & \nodata & \nodata && 0.65$^{+0.07}_{-0.06}$ & 692.1$^{+111.8}_{-104.4}$ & -10.56 && 10.67$^{+1.36}_{-1.15}$ & 396.1$^{+25.1}_{-25.3}$ & -10.16 & $<$0.01 & 1.48 & 30.95 \\
    28946 & E & \nodata & \nodata & \nodata && 0.68$^{+0.08}_{-0.07}$ & 717.6$^{+112.1}_{-111.0}$ & -10.53 && 10.73$^{+1.70}_{-1.37}$ & 410.8$^{+34.6}_{-35.2}$ & -10.14 & $<$0.01 & 1.57 & 30.97 \\
    28946 & R & \nodata & \nodata & \nodata && 0.68$^{+0.07}_{-0.06}$ & 785.9$^{+112.1}_{-102.4}$ & -10.49 && 10.08$^{+1.46}_{-1.20}$ & 395.5$^{+35.7}_{-35.6}$ & -10.16 & $<$0.01 & 1.44 & 30.97 \\
    61421 & Q & 0.11$\pm$0.01 & 135.7$\pm$48.8 & -11.68 && 0.25$^{+0.02}_{-0.01}$ & 24.7$^{+13.4}_{-13.0}$ & -11.88 && 0.77$^{+0.25}_{-0.16}$ & 0.3$\pm$0.1 & -13.18 & 0.29 & 2.17 & 27.70 \\
    63433 & Q & \nodata & \nodata & \nodata && 0.42$^{+0.06}_{-0.05}$ & 10.8$^{+1.0}_{-1.1}$ & -12.14 && 0.87$^{+0.11}_{-0.09}$ & 0.8$\pm$0.3 & -12.58  & 0.17 & 0.93 & 28.78 \\
    82443 & Q & 0.32$\pm$0.12 & 16.7$^{+12.6}_{-12.5}$ & -11.62 && 0.81$\pm$0.22 & 11.4$^{+7.1}_{-7.3}$ & -11.73 && 1.50$^{+0.58}_{-0.54}$ & 6.7$\pm$5.6 & -11.99 & 1 & 1.23 & 29.27 \\
    82443 & E & 0.32$\pm$0.02 & 43.0$\pm$8.9 & -11.46 && 0.85$\pm$0.05 & 21.5$\pm$3.0 & -11.57 && 3.31$^{+1.68}_{-1.06}$ & 2.7$\pm$0.9 & -12.45 & 0.28 & 1.31 & 29.40 \\
    82443 & F & 0.49$\pm$0.06 & 9.9$\pm$3.8 & -11.53 &&	0.97$^{+0.12}_{-0.10}$ & 6.0$\pm$2.4 & -11.76 && 4.08$^{+2.71}_{-1.36}$ & 7.2$\pm$1.5 & -11.88 & 1 & 1.27 & 29.37 \\
    82558 & Q & 0.28$\pm$0.02 & 80.0$^{+9.2}_{-9.3}$ & -11.37 && 0.97$\pm$0.06 & 92.2$^{+20.8}_{-20.9}$ & -11.05 && \nodata & \nodata & \nodata & 0.18 & 1.56 & 29.74 \\
    82558 & D & 0.28$^{+0.01}_{-0.02}$ & 87.9$^{+6.6}_{-7.3}$ & -11.34 && 0.96$^{+0.03}_{-0.05}$ & 94.5$^{+13.4}_{-16.0}$ & -11.05 && \nodata & \nodata & \nodata & 0.17 & 2.08 & 29.77 \\
    82558 & F & 0.28$\pm$0.01 & 85.8$^{+9.8}_{-9.9}$ & -11.32 && 0.97$\pm$0.03 & 90.9$^{+6.2}_{-6.1}$ & -11.05 && 2.18$^{+0.60}_{-0.47}$ & 35.5$^{+16.1}_{-16.2}$ & -11.43 & 0.19 & 1.78 & 29.84 \\
    82558 & E & 0.29$\pm$0.01 & 89.6$^{+9.6}_{-9.7}$ & -11.35 && 0.98$\pm$0.03 & 105.2$^{+21.5}_{-21.8}$ & -11.02 && 2.15$^{+1.84}_{-0.67}$ & 16.0$^{+11.4}_{-8.2}$ & -11.77 & 0.16 & 1.53 & 29.77 \\
    120136 & Q & 0.10$\pm$0.02 & 13.5$^{+15.4}_{-7.8}$ & -12.48 && 0.41$^{+0.11}_{-0.12}$ & 11.2$^{+6.8}_{-6.5}$ & -11.89 && 0.70$^{+0.11}_{-0.10}$ & 3.8$^{+4.4}_{-3.2}$ & -12.22 & 1 & 1.31 & 28.76 \\
    129333 & Q & \nodata & \nodata & \nodata && 0.50$^{+0.07}_{-0.06}$ & 31.2$^{+4.4}_{-3.9}$ & -11.57 && 0.98$^{+0.06}_{-0.05}$ & 45.8$^{+5.0}_{-5.8}$ & -11.34 & 0.20 & 1.24 & 30.01 \\
    129333 & E & \nodata & \nodata & \nodata && 0.53$^{+0.09}_{-0.08}$ & 32.2$^{+6.5}_{-5.0}$ & -11.56 && 0.98$^{+0.07}_{-0.05}$ & 50.7$^{+5.2}_{-7.0}$ & -11.30 & 0.19 & 1.70 & 30.04 \\
    129333 & D & \nodata & \nodata & \nodata && 0.52$\pm$0.09 & 33.8$^{+9.1}_{-8.6}$ & -11.54 && 1.00$\pm$0.08 & 53.1$^{+11.9}_{-12.2}$ & -11.28 & 0.20 & 1.84 & 30.06 \\
    129333 & R & \nodata & \nodata & \nodata && 0.54$^{+0.07}_{-0.06}$ & 32.8$^{+4.6}_{-3.2}$ & -11.52  && 1.06$^{+0.06}_{-0.04}$ & 49.2$^{+3.7}_{-5.3}$ & -11.31 & 0.21 & 2.00 & 30.05 \\
    129333 & F & 3.01$^{+0.57}_{-0.36}$ & 42.4$^{+5.4}_{-6.5}$ & -11.27 && 0.45$^{+0.05}_{-0.03}$ & 33.4$^{+3.0}_{-2.9}$ & -11.52 && 1.01$^{+0.05}_{-0.03}$ & 55.0$^{+4.1}_{-3.9}$ & -11.23 & 0.24 & 1.55 & 30.31 \\
    162020 & Q & 0.27$\pm$0.03 & 5.1$\pm$0.6 & -12.5 && 0.92$\pm$0.07 & 3.5$\pm$0.5 & -12.38 && \nodata & \nodata & \nodata & 0.23 & 1.16 & 28.92 \\
    165567 & Q & 0.13$^{+0.08}_{-0.03}$ & 0.5$^{+0.5}_{-0.3}$ & -13.25 && 0.73$\pm$0.03 & 1.7$\pm$0.1 & -12.26 && \nodata & \nodata & \nodata & 1 & 1.44 & 29.30 \\
    179949 & Q & 0.12$\pm$0.04 & 1.4$^{+2.3}_{-1.0}$ & -13.16 && 0.53$\pm$0.09 & 0.7$\pm$0.2 & -12.65 && 0.60$^{+0.09}_{-0.07}$ & 0.5$\pm$0.2 & -12.86 & 1 & 1.43 & 28.42 \\
    189733 & Q & 0.29$\pm$0.1 & 1.8$^{+2.2}_{-2.3}$ & -12.82 && 0.82$^{+0.19}_{-0.16}$ & 0.5$^{+0.4}_{-0.3}$ & -12.95 && \nodata & \nodata & \nodata & 1 & 1.33 & 28.06 \\
    189733 & E & 0.28$^{+0.10}_{-0.12}$ & 0.7$^{+0.3}_{-0.2}$ & -12.82 && 0.97$^{+0.30}_{-0.26}$ & 0.5$^{+0.1}_{-0.2}$ & -12.91 && \nodata & \nodata & \nodata & 1 & 1.45 & 28.11 \\
    190771 & Q & 0.10$\pm$0.02 & 5.3$^{+5.4}_{-2.3}$ & -12.54 && 0.45$^{+0.17}_{-0.09}$ & 2.8$^{+1.4}_{-1.3}$ & -12.10 && 0.75$^{+0.13}_{-0.07}$ & 2.8$^{+1.2}_{-1.7}$ & -12.03 & 1 & 1.47 & 28.94 \\
    190771 & R & 0.10$\pm$0.04 & 3.1$^{+37.3}_{-2.2}$ & -12.83 && 0.45$^{+0.08}_{-0.05}$ & 4.4$\pm$0.5 & -11.90 && 0.93$^{+0.08}_{-0.06}$ & 3.1$^{+0.05}_{-0.06}$ & -12.04 & 1 & 1.36 & 29.00 \\
    190771 & E & \nodata				& \nodata		& \nodata && 0.52$\pm$0.05 & 12.9$^{+2.5}_{-1.9}$ & -11.80 && 0.94$^{+0.06}_{-0.05}$ & 11.5$^{+2.1}_{-1.9}$ & -11.81 & 0.33 & 1.21 & 29.13 \\
    190771 & F & 0.10$\pm$0.02 & 6.8$^{+8.6}_{-3.6}$ & -12.67 && 0.61$^{+0.03}_{-0.04}$ & 10.0$^{+2.8}_{-2.7}$ & -11.70 && 1.09$^{+0.08}_{-0.07}$ & 6.2$^{+1.9}_{-1.6}$ & -11.98 & 0.56 & 1.19 & 29.15 \\
    192020 & Q & 0.09$^{+0.02}_{-0.03}$ & 2.3$^{+7.5}_{-1.1}$ & -13.31 && 0.32$^{+0.05}_{-0.04}$ & 0.9$\pm$0.2 & -12.93 && 0.88$^{+0.08}_{-0.07}$ & 0.6$\pm$0.1 & -12.99 & 0.49 & 1.28 & 28.28 \\
    \enddata

\tablenotetext{\dag}{The global APEC abundance parameter, $Z$, in solar units. Values of $Z=1$ were fixed at solar abundance; non-unity values indicate cases where $Z$ was left a free parameter.}
\end{deluxetable*}
\vspace{-0.85cm}
Individual stars in our sample required different numbers of thermal plasma components to obtain statistically acceptable fits. Figure~\ref{fig:kT_distribution} shows a histogram of all best-fit APEC temperature components from the quiescent spectral models: stars fit with 1T, 2T, and 3T models therefore contribute one, two, and three temperatures, respectively. The resulting distribution shows preferred fitted component temperatures near $kT\sim$0.1 keV ($\sim$1.2 MK), $\sim$0.4 keV ($\sim$4.6 MK), and $\sim$0.8 keV ($\sim$9.3 MK). We interpret these as characteristic temperatures recovered by a low-order, multi-temperature APEC parameterization of intrinsically multi-thermal coronal emission, rather than as evidence for physically discrete isothermal plasma sources. The recovered component temperatures may also be influenced by the line emissivities that dominate the spectra and by the instrumental response. Generally, stars closer to our net count cutoff ($\sim$2000 net counts for \xmm\ observations and $\sim$500 net counts for \chandra\ observations) did not require the lowest temperature component, as \xmm\ and especially \chandra\ do not have the soft X-ray sensitivity necessary to constrain this component in fainter stars. On the other hand, the lack of a hottest component in some stars is likely real, as both telescopes have sufficient sensitivity to these harder photons if they are present.

\begin{figure}
    \centering
    \includegraphics[width=1\linewidth]{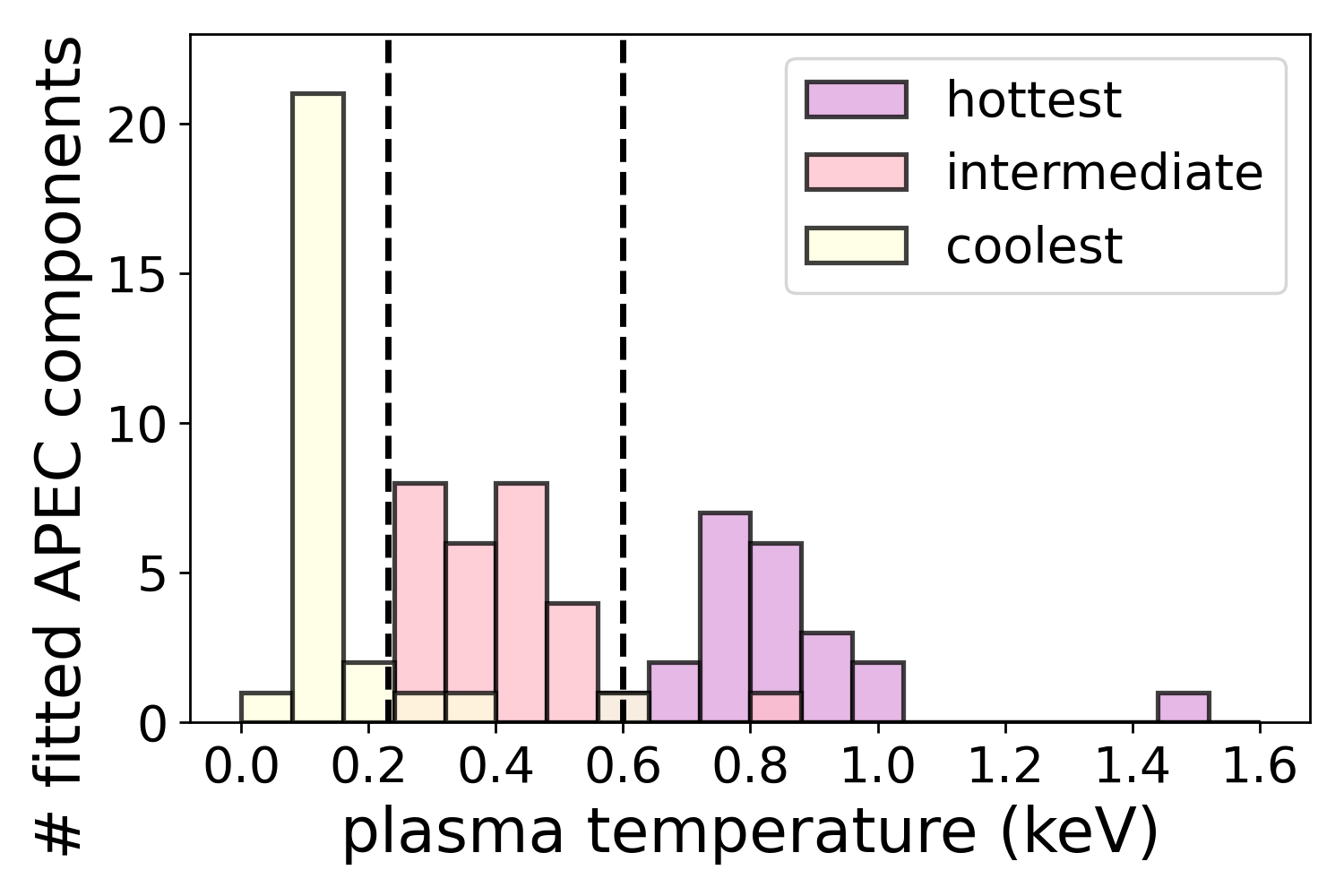}
    \caption{Distribution of best-fit APEC temperature components from the quiescent spectral fits from Table~\ref{tab:avg_spectra}. Each fitted temperature component is included separately, so stars modeled with 1T, 2T, and 3T APEC models contribute one, two, and three entries, respectively. The colored histograms indicate approximate cool (light yellow), intermediate (pink), and hot (purple) fitted-temperature groupings. Vertical dashed lines correspond to approximate boundaries between these fitted-temperature groupings at $\sim$0.24 keV and $\sim$0.60 keV.}
    \label{fig:kT_distribution}
\end{figure}

The boundaries between these fitted-temperature groupings can be approximately drawn at $\sim$0.24 keV and $\sim$0.60 keV, which corresponds to temperatures of $\sim$2.7 MK and $\sim$7.0 MK, respectively (i.e., the vertical dashed lines in Figure~\ref{fig:kT_distribution}). As part of our spectral modeling we used the \texttt{cflux} convolution model to estimate the 0.3-10 keV flux produced by each thermal plasma component. We then calculated the fraction of the overall 0.3-10 keV flux that was produced by the coolest and hottest components. The stars with higher surface fluxes have the largest flux contribution from the hottest plasma component; for stars with \Fx$\gtrsim10^6$ \flux, roughly half of the total 0.3-10 keV flux is originating from material hotter than 7.0 MK. The inverse is also true: stars with the lowest surface flux have their X-ray emission dominated by the coolest plasma components. The correlations between X-ray surface flux and the flux fraction from cool and hot components is shown in Figure~\ref{fig:fluxfrac_temp}

\begin{figure}
    \centering
    \includegraphics[width=1\linewidth]{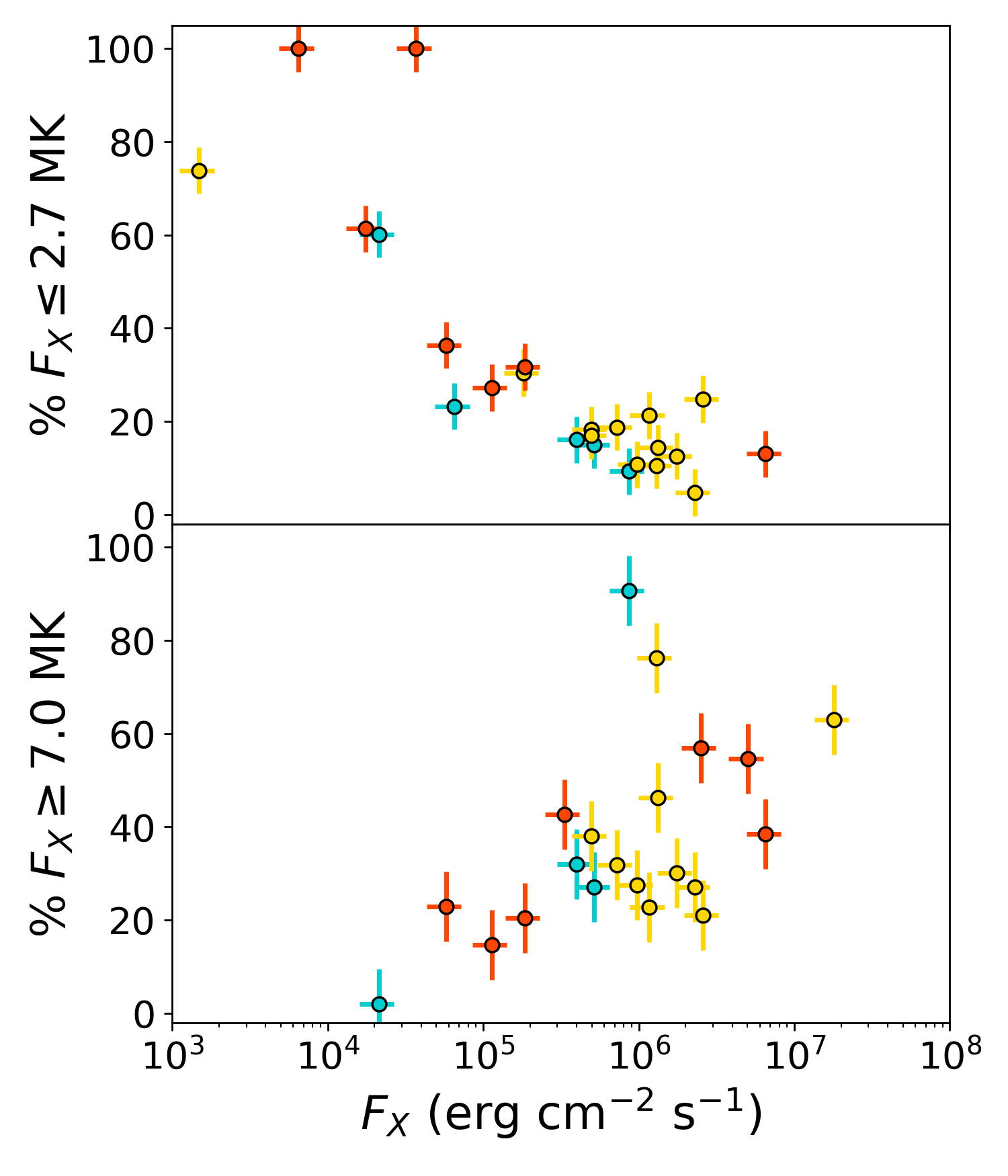}
    \caption{The flux fraction originating from the coolest ($\leq$2.7 MK) thermal plasma component ({\it top}) and the hottest ($\geq$7.0 MK) thermal plasma component ({\it bottom}) as a function of X-ray surface flux. Stars are color coded as in Figure~\ref{fig:Age_dist}.}
    \label{fig:fluxfrac_temp}
\end{figure}

We used the best-fit normalizations on each model component, which are related to the emission measure, to compute a single, emission measure-weighted coronal temperature $T_{\rm EM}$, defined as
\vspace{-0.1cm}
\begin{equation}
    \text{log}T_{\rm EM}=\frac{ \sum_i EM_i \text{log}T_i}{\sum_i EM_i},
\end{equation}

\noindent where $T_i$ (in K) and $EM_i$ are the temperature and emission measure of each APEC component. This definition reduces the leverage of the hottest component relative to a linear-temperature weighted mean \citep{Johnstone+15,Johnstone2021}.

Figure 6 shows the distribution of $T_{\rm EM}$ and their relationships with X-ray surface flux \Fx\ and luminosity \Lx. \citet{Johnstone+15} found that coronal temperature correlates more tightly with \Fx\ than with \Lx\ or with \Lx/\Lbol, consistent with \Fx\ being a more direct tracer of magnetic heating per unit stellar surface area. We therefore fit the $T_{\rm EM}$-\Fx\ relation for our sample using a power law model and find a best-fit relation
\vspace{-0.1cm}
\begin{equation}\label{eq:TEM-Fx}
    T_{\rm EM}=\left(2.0\pm0.3 \right) \left(\frac{F_X}{10^5~\text{erg~cm}^{-2}~\text{s}^{-1}} \right)^{0.24\pm0.05} \text{MK}.
\end{equation}

\noindent Our best-fit exponent, 0.24$\pm$0.05, is consistent with the $T\propto F_X^{0.26}$ scaling reported by \citet{Wood+18} and \citet{Johnstone+15}. For comparison, we also fit a similar power law relationship to $T_{\rm EM}$ and \Lx\ and find:
\vspace{-0.1cm}
\begin{equation}\label{eq:TEM-Lx}
    T_{\rm EM}=\left(0.62^{+0.25}_{-0.18} \right) \left(\frac{L_X}{10^{26}~\text{erg}~\text{s}^{-1}} \right)^{0.27\pm0.05} \text{MK}.
\end{equation}

\noindent The $T_{\rm EM}$-\Fx\ relation is broadly consistent with the expectation that coronal temperature increases with increasing magnetic activity and provides the most direct comparison to the surface-flux scaling reported by \citet{Johnstone+15}. The $T_{\rm EM}$-\Lx\ relation shows a similar trend but includes the additional dependence of luminosity on stellar radius.

\begin{figure*}
    \centering
    \begin{tabular}{ccc}
        \includegraphics[width=0.315\linewidth]{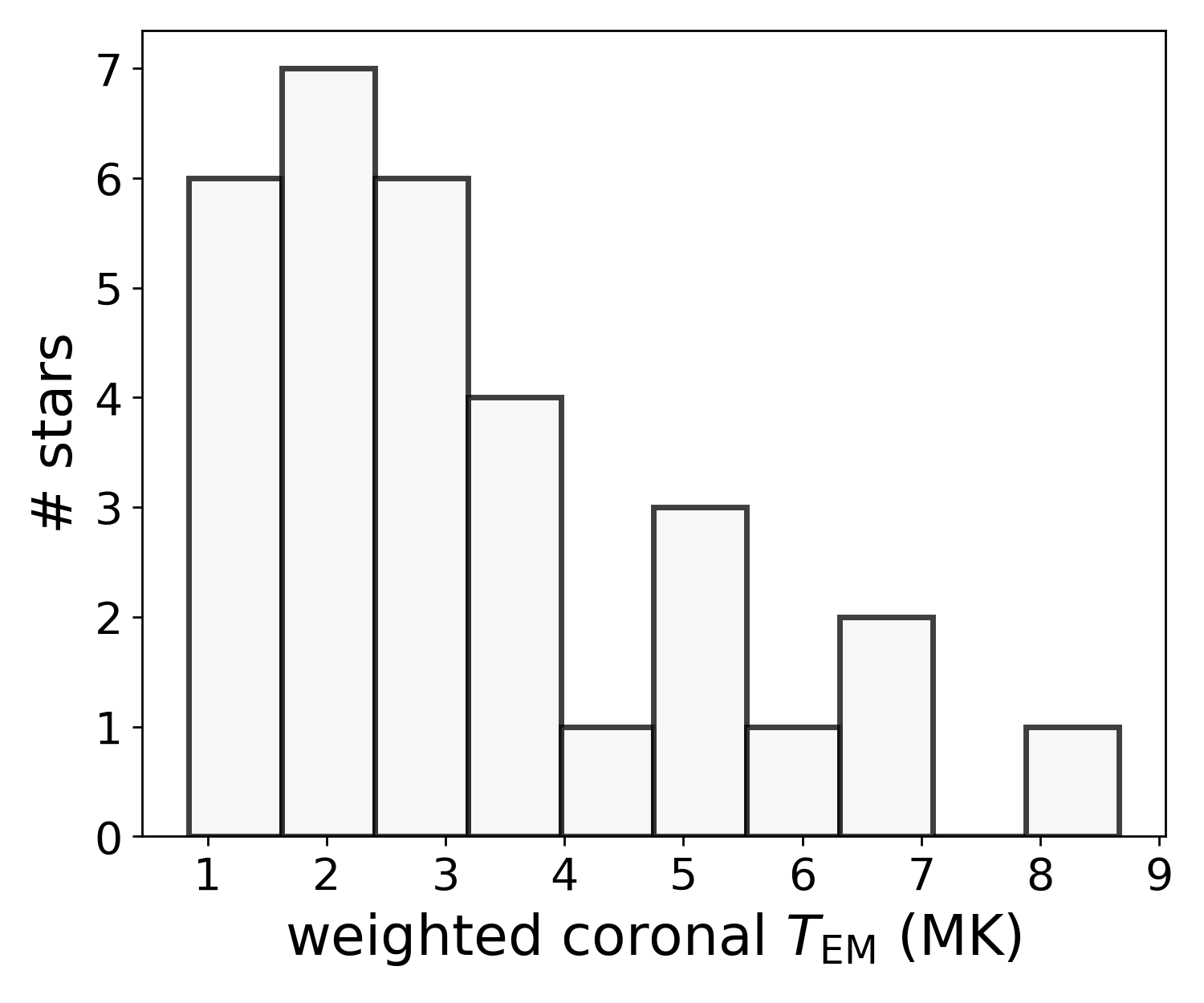} &
        \includegraphics[width=0.315\linewidth]{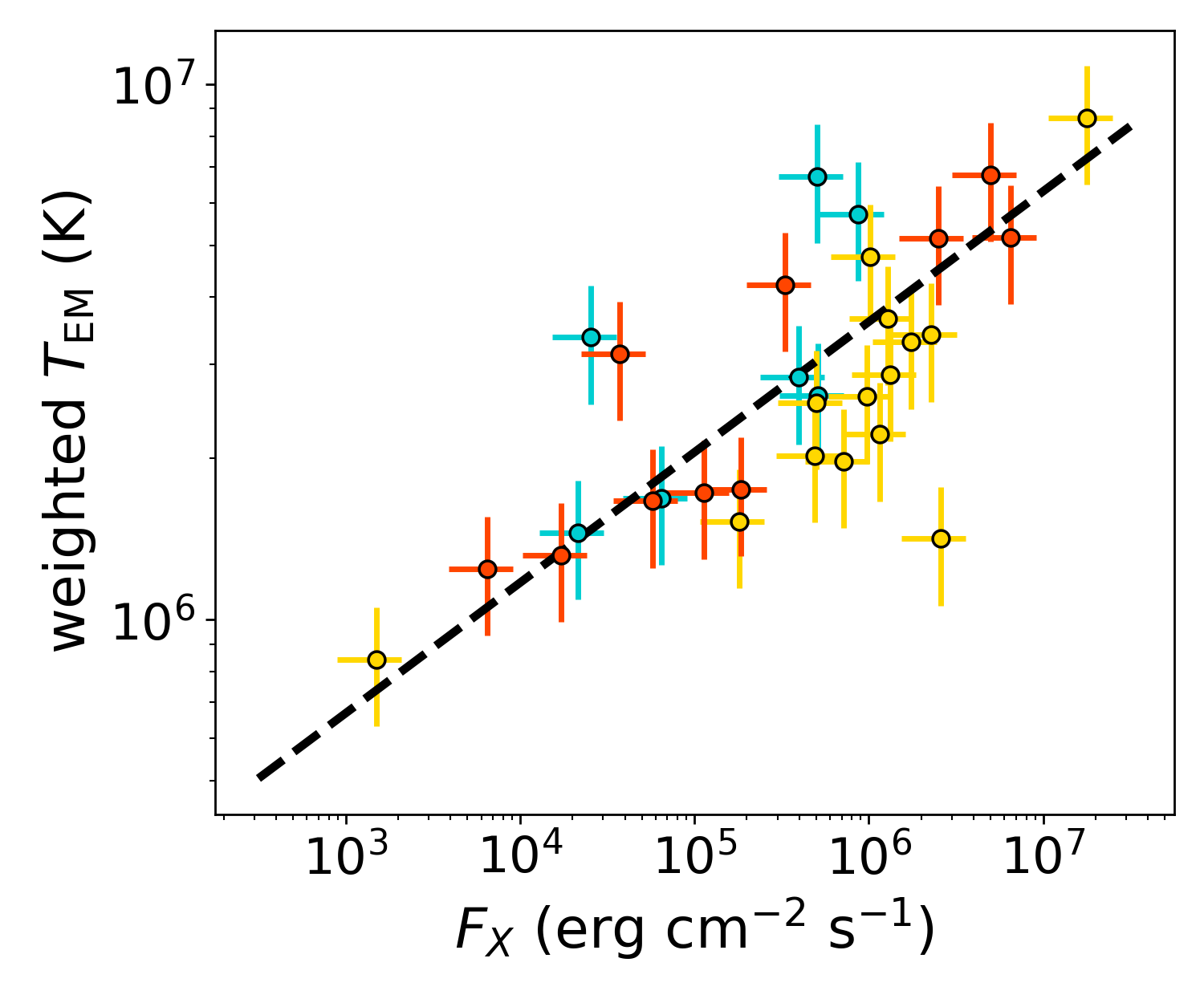} &
        \includegraphics[width=0.315\linewidth]{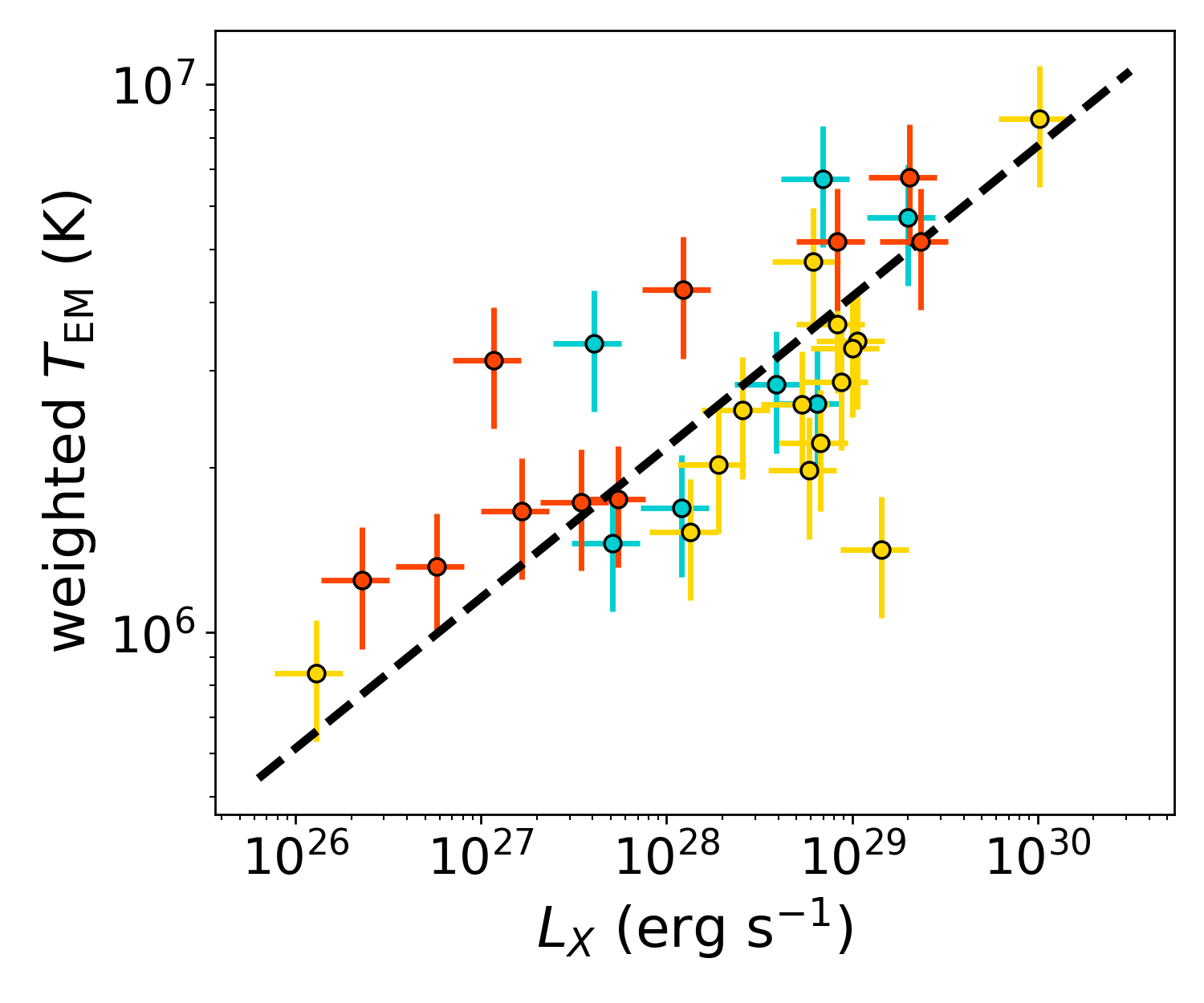} \\
    \end{tabular}
    \caption{{\it Left}: The distribution of emission measure-weighted coronal temperatures $T_{\rm EM}$ (in MK) for all stars in our sample. {\it Middle}: The relationship between $T_{\rm EM}$ and surface 0.3-10 keV flux \Fx. {\it Right}: The relationship between $T_{\rm EM}$ and 0.3-10 keV \Lx\ of the star. The black dashed lines in the middle and right panels show the best fit relationships given by Equation~\ref{eq:TEM-Fx} (middle panel) and Equation~\ref{eq:TEM-Lx} (right panel), discussed in the text. Stars are colored by spectral type as in Figure~\ref{fig:Age_dist}. }
    \label{fig:coronalT_trends}
\end{figure*}

The relationship between \Lx\ and $T_{\rm EM}$ enables the selection of an appropriate plasma temperature for converting the observed count rates from stars too faint for direct spectral modeling. While some of these low-count stars are relatively bright, but not observed with long enough exposure times for enough counts to be collected, most stars for which we are unable to perform spectral modeling are intrinsically fainter. For bright stars with short X-ray exposures, we assume an APEC model with $kT$=0.4 keV to convert the observed \xmm\ or \chandra\ count rates to 0.3-10 keV fluxes. For intrinsically fainter sources, we generally select $kT\sim$0.1-0.2 keV to estimate 0.3-10 keV fluxes.

In Table~\ref{tab:avg_Lx}, we provide X-ray variability metrics, median X-ray luminosities, and \Lx/\Lbol\ ratios for all stars in our present sample. For stars with spectral modeling, we report only the quiescent \Lx\ in Table~\ref{tab:avg_Lx}. Variability metrics ($A^2$ and reduced $\chi^2$) are reported for all stars bright enough for a light curve to be extracted; for stars with multiple observations, we report the maximum values of $A^2$ and $\chi^2$ measured across all light curves. We additionally compare the observed \Lx/\Lbol\ ratios of our sample stars to the range of the modern-day Sun \citep[$\sim$(2-25)$\times10^{-6}$;][]{Linsky+20} and stars with average quiescent \Lx/\Lbol\ values that fall within this typical Solar range are labeled as ``Sun-like'' in Table~\ref{tab:avg_Lx}.

\begin{deluxetable}{cccccc}
\tabletypesize{\footnotesize}
    \tablecaption{Median X-ray Luminosities Compared to Solar}\label{tab:avg_Lx}
\tablehead{ 
                 & \multicolumn{2}{c}{variability metrics} & &  \\ \cline{2-3}
    \colhead{HD} & max $A^2$ & max $\chi^2$ & log\Lx$^*$ & log(\Lx/\Lbol) & Sun-Like?
 }
    \startdata 
    1461 & 18.52 & 0.5 & 27.24 & -6.43 & \checkmark  \\
    1835 & 0.54 & 1.6 & 28.92 & -4.70 &  \\
    3651 &       &     & 26.35 & -6.95 &  \\
    3765 &       &     & 26.61 & -6.55 & \checkmark  \\
    8673 & 3.36 & 1.7 & 28.85 & -5.26 &  \\
    11505 & 8.32 & 10.3 & 28.21 & -5.52 &   \\
    25680 &	0.72 & 2.0 & 28.33 & -5.27 &   \\
    25874 &      &      & $<$27.58 & $<$-6.02 \\
    27442 &      &      & $<$26.45 & $<$-7.98 \\
    28099 &	0.95 & 1.2  & 28.31 & -5.29 & \checkmark  \\
    28946 & 7.06 & 5.9 & 30.95 & -2.32 &  \\
    33636 & 3.25 & 0.6  & 27.43 & -6.21 & \checkmark  \\
    38529 & 0.53 & 1.2 & 28.56 & -5.85 & \checkmark  \\
    61421 & 1.61 & 1.5 & 27.70 & -6.73 & \checkmark  \\
    63433 & 0.50 & 6.5 & 28.78 & -4.68 &  \\
    75289 &      &     & 27.24 & -6.65 & \checkmark  \\
    79969 & 3.54 & 1.5 & 27.16 & -6.09 & \checkmark  \\
    82443 & 7.47 & 1.8 & 29.27 & -3.98 &  \\
    82558 & 4.80 & 11.1 & 29.74 & -5.26 &  \\
    99491 &	1.37 & 2.1 & 27.63 & -5.75 & \checkmark   \\
    102117 &     &     & $<$28.81 & $<$-4.93 \\
    108147 & 6.73 & 0.8 & 27.65 & -6.23 & \checkmark  \\
    108309 & 13.78 & 0.5 & 27.15 & -6.72 & \checkmark  \\
    114386 &      &     & $<$27.32 & $<$-5.72 \\
    114783 &      &     & $<$26.24 & $<$-6.96 \\
    120066 &      &     & $<$27.85 & $<$-6.12 \\
    120136 & 0.86 & 1.6 & 28.76 & -5.32 &   \\
    120690 & 0.66 & 2.2 & 27.78 & -5.71 & \checkmark  \\
    129333 & 43.49 & 9.2 & 30.01 & -3.53 &  \\
    130307 & 1.22 & 1.6 & 27.37 & -5.70 & \checkmark  \\
    135599 & 0.28 & 1.1 & 27.84 & -5.37 & \checkmark  \\
    145417 &      &     & $<$27.01 & $<$-6.02 \\
    150248 &      &     & $<$26.38 & $<$-7.23 \\
    157214 &      &     & 27.13 & -6.56 & \checkmark  \\
    157347 &      &     & $<$26.67 & $<$-6.94 \\
    162020 & 1.05 & 1.7 & 28.92 & -4.07 &  \\
    165567 & 1.18 & 1.9 & 29.30 & -5.02 &  \\
    170657 & 3.89 & 2.1 & 27.05 & -6.07 & \checkmark  \\
    179949 & 3.46 & 2.5 & 28.42 & -5.46 &  \\
    187237 & 8.37 & 1.7 & 27.57 & -6.05 & \checkmark  \\
    189733 & 4.05 & 1.5 & 28.06 & -5.04 &  \\
    190771 & 1.07 & 7.5 & 28.94 & -4.67 &  \\
    192020 & 2.85 & 2.4 & 28.28 & -4.92 &  \\	
    193664 &      &     & $<$27.43 & $<$-6.22 \\
    207740 &      &     & 28.01 & -5.64 & \checkmark  \\
    209779 &      &     & 28.04 & -5.60 &   \\
    210918 &      &     & $<$26.18 & $<$-7.55 \\
    217107 &      &     & $<$26.72 & $<$-6.96 \\
    \enddata
\tablecomments{$^*$Where \Lx\ is in units of \lum\ and measured in the 0.3-10 keV energy range.}
\end{deluxetable}

\vspace{-0.35cm}
\section{Age Dependence of X-ray Emission}\label{sec:age_dependence}
In both this work and in \citet{Binder+24}, the 0.3-10 keV energy range is adopted as the ``default'' energy band for measuring quantities such as \Fx\ and \Lx\ because it is energy range over which \xmm\ is most sensitive (the preferred \chandra\ energy range is typically 0.5-7 keV, with extrapolations to the wider 0.3-10 keV made using the best-fit spectral models within XSPEC). However, much previous work -- and most of the observable X-ray emission from stellar coronae -- utilized the 0.1-2.4 keV ROSAT band. Converting between these energy bands requires knowledge of the underlying spectral shape, which is set by the temperature of the plasma.

Using the APEC model, we generate a grid of 100 thermal plasmas of varying temperature, logarithmically spaced from the minimum observed $kT$ in our spectral modeling ($\sim$0.08 keV) up to $\sim$2 keV. At each temperature, we calculate: the fraction of the {\it full} 0.1-10 keV flux that is emitted in the ROSAT band (0.1-2.4 keV) and the \xmm\ band (0.3-10 keV), as well as the fraction of the 0.1-2.4 keV flux to the 0.3-10 keV flux. These flux ratios are shown in Figure~\ref{fig:energy_band_conversions}. The 0.1-2.4 keV flux makes up $>$90\% of the full X-ray flux for thermal plasmas below $\sim$1 keV in temperature. The 0.3-10 keV band, meanwhile, captures $\sim$80\% of the total flux for plasmas above $\sim$0.25 keV, and drops precipitously (to $\sim$15\%) at cooler temperatures. Only above plasma temperatures of $\sim$1.5 keV does the 0.3-10 keV flux exceed the 0.1-2.4 keV flux. As a result, the 0.1-2.4 keV to 0.3-10 keV flux ratio is nearly constant (at $\sim$1.2) for temperatures $\gtrsim$0.25 keV and increases dramatically for cooler plasmas. We compare the model predicted 0.1-2.4 keV to 0.3-10 keV flux ratio as a function of emission measure-weighted coronal temperature to the best-fit quiescent spectra summarized in Table~\ref{tab:avg_Lx}. The biggest discrepancy between the simple, single-temperature APEC model and the multi-temperature models used to describe the data occurs for temperatures of $\sim$0.2-0.4 keV, where the data show flux ratios $\sim$40\% higher than the single-temperature model prediction. We use the modeled ratio of the 0.1-2.4 keV flux to the 0.3-10 keV flux, as a function of temperature, to estimate the 0.1-2.4 keV ROSAT-band \Lx\ from the harder, 0.3-10 keV \xmm\ band for comparison to literature results.

\begin{figure}
    \centering
    \includegraphics[width=1\linewidth]{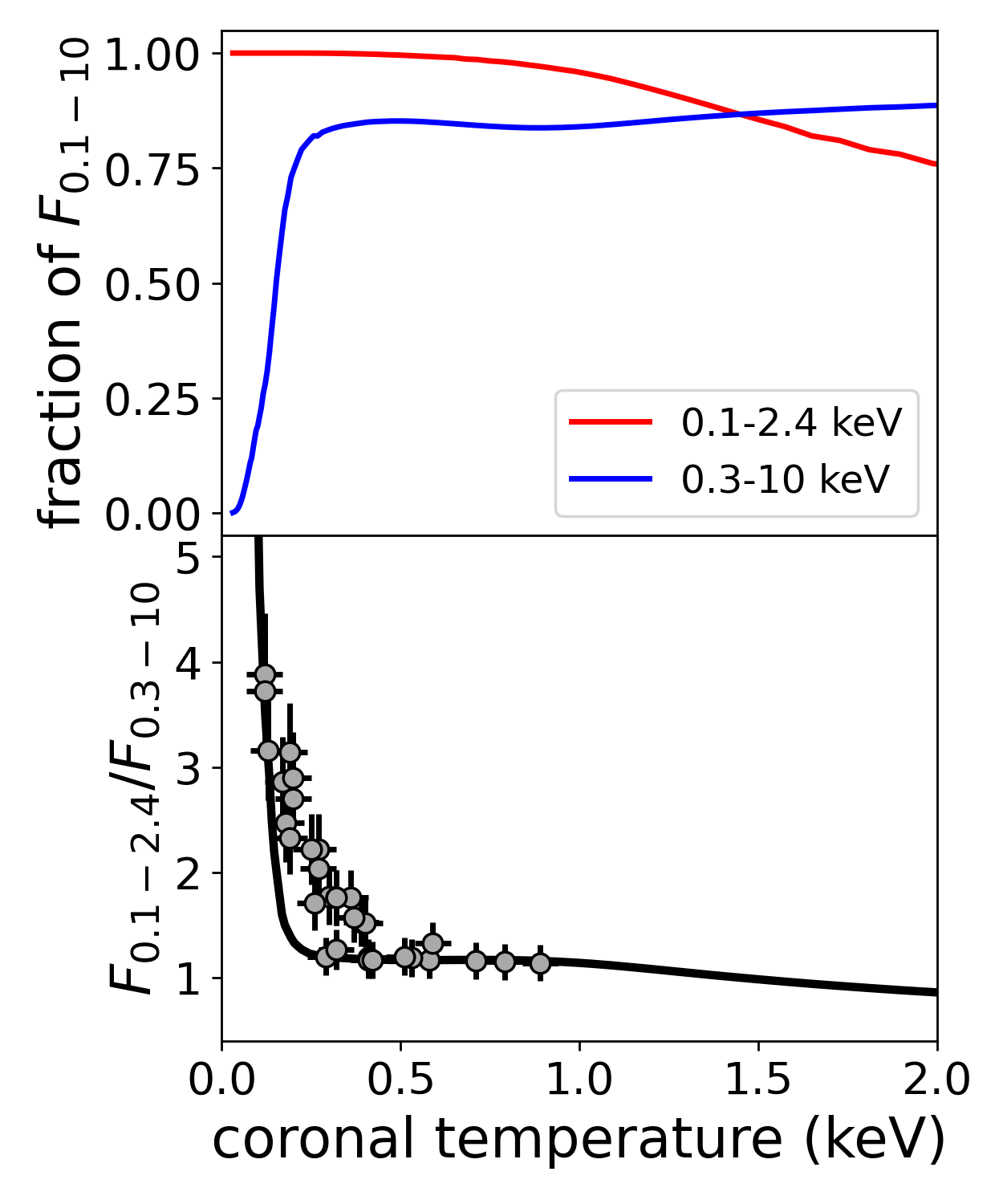}
    \caption{{\it Top:} The fraction of the 0.1-10 keV flux emitted in the soft, 0.1-2.4 keV ROSAT band (red) and the harder 0.3-10 keV \xmm\ band (blue), as a function of coronal temperature for an \textsc{apec} thermal plasma model. {\it Bottom}: The ratio of the soft 0.1-2.4 keV ROSAT band to the harder 0.3-10 keV \xmm\ band as a function of coronal temperature. The gray circles represent the observed ratios calculated from the best-fit, multi-temperature quiescent spectra in Table~\ref{tab:avg_spectra}.}
    \label{fig:energy_band_conversions}
\end{figure}

Figure~\ref{fig:Lx_age} shows the age dependence of X-ray emission for the stars in our sample \citep[inclusive of the][ sample]{Binder+24}, using both the measured 0.3-10 keV values (on the left side) and the inferred 0.1-2.4 keV values (on the right side). In both panels, Equation~\ref{eq:Gudel04} is shown by a dashed black line (top panels). In the middle row of the figure, we show the \Fx-age relationships in both bandpasses, and in the bottom row we plot \Fx/\vsini\ as a function of age in both bandpasses. We adopt the Solar radius $R_{\odot}$ and equatorial rotation velocity ($\sim$2 km s$^{-1}$) to express the \citet{Gudel04} \Lx\--age relation as a solar-reference \Fx/$v$--age envelope for comparison with the middle and lower panels of Figure~\ref{fig:Lx_age}. The average value and range in values exhibited by the Sun are also shown for reference. In general, the inferred 0.1-2.4 keV luminosities of the stars in our sample are in good agreement with \citet{Gudel04}, while the 0.3-10 keV luminosities show considerably more scatter at ages $\gtrsim$4 Gyr. This is likely the result of the rapid decay in the hottest component of the corona (which the high energy \xmm\ band is most sensitive to) with age, as stellar rotation rates slow and magnetic heating becomes less efficient. We also note that low-activity stars can show large fractional variability because their X-ray emission may be dominated by only a few active regions, so rotational modulation and active-region evolution can contribute to the observed scatter, as in the Sun.

\begin{figure*}
    \centering
    \begin{tabular}{cc}
        \includegraphics[width=0.49\linewidth]{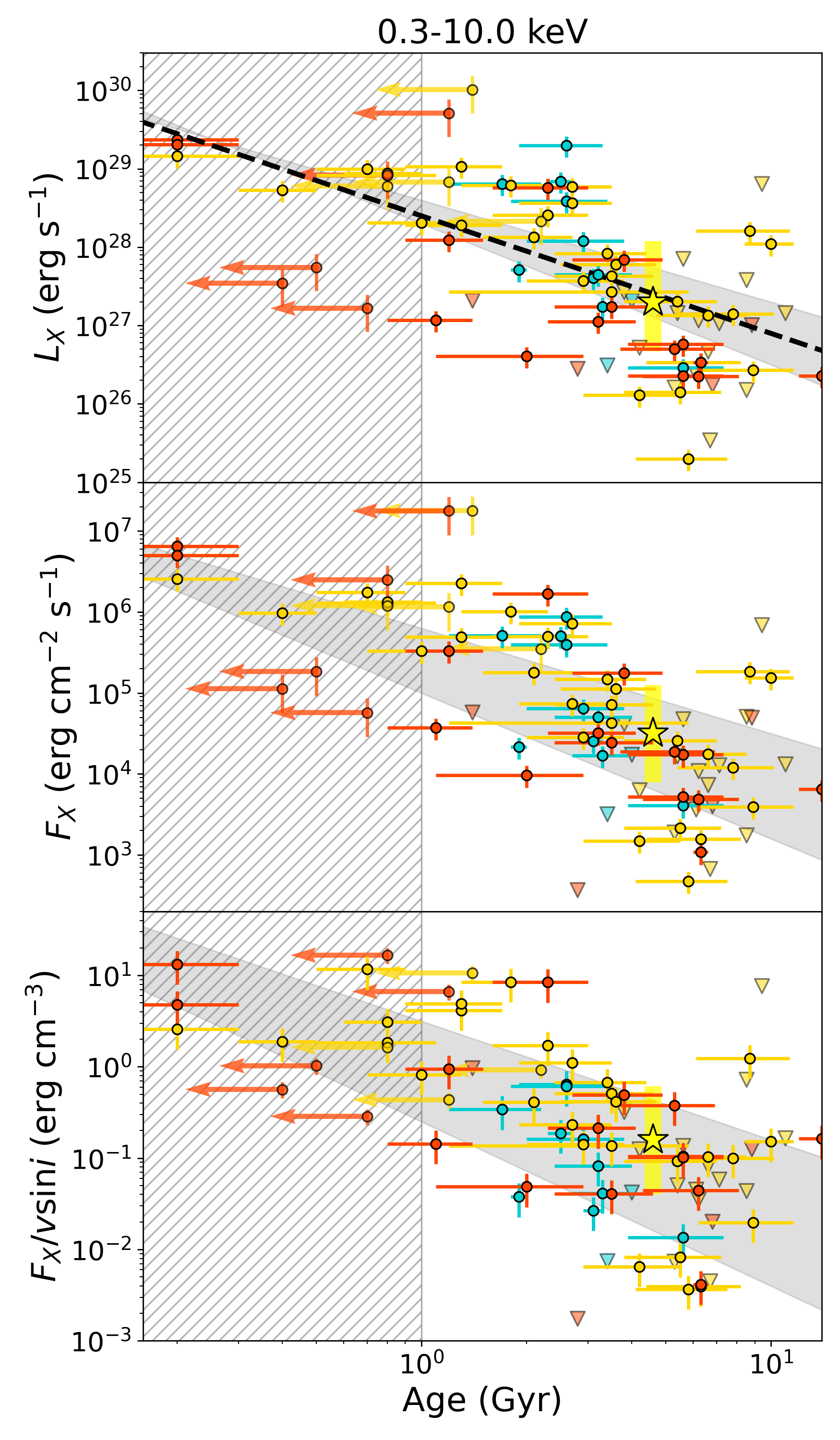} &
        \includegraphics[width=0.49\linewidth]{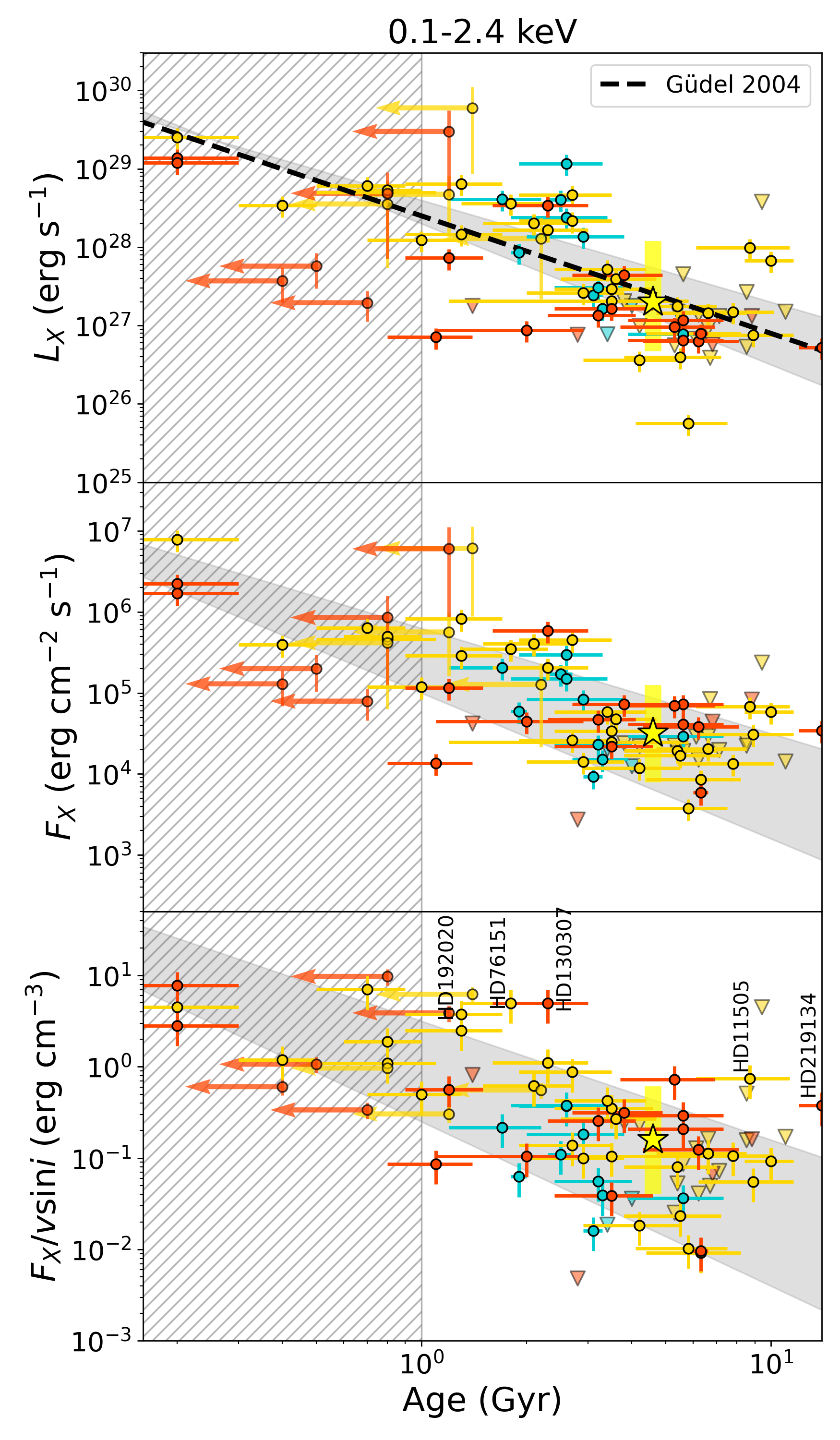} \\
    \end{tabular}
    \caption{The evolution of \Lx\ ({\it top} panels), \Fx\ ({\it middle} panels), and \Fx/\vsini\ with age ({\it bottom} panels). The dashed black line shows the measured \Lx-age relationship from \citet[][see Equation~\ref{eq:Gudel04}]{Gudel04}. {\it Left} panels are constructed using the observed 0.3-10 keV \xmm\ energy band. {\it Right} panels show the inferred 0.1-2.4 keV ROSAT energy band (using the black curve from the bottom panel of Figure~\ref{fig:fluxfrac_temp}). The gray shaded region shows the \citet{Gudel04} \Lx-age uncertainty envelope, scaled as a solar-reference \Fx-age and \Fx/$v$-age envelope for comparison in the middle and lower panels. In all panels, the region of the diagrams occupied by the Sun is shown by the yellow star and yellow shaded region. Stars with X-ray upper limits (non-detections) are shown with downward-pointing triangles and stars with age upper limits are indicated with left-pointing arrows. The gray hatched region indicates ages $<$1 Gyr. Stars with \Fx/$v$sin$i\geq1\sigma$ above the upper boundary of the gray reference envelope in the lower-right panel are labeled.}
    \label{fig:Lx_age}
\end{figure*}

%\vspace{0.5cm}
\subsection{Outlier Stars}\label{sec:outliers}
We next consider stars with elevated activity relative to the reference trends in Figure~\ref{fig:Lx_age}. We caution that the gray envelopes shown in Figure~\ref{fig:Lx_age} should not be interpreted as a unique age predictor for stars younger than $\sim$1 Gyr due to non-unique rotational evolution, which produces a broad range of activity levels in this regime; this region of the figure is shown with a gray-hatched background. We do not interpret young stars outside the reference envelope as anomalous in the same sense as older stars. For the case-by-case discussion below, we focus on stars that lie more than $1\sigma$ above the upper boundary of the gray reference envelope in the 0.1-2.4 keV \Fx/$v$sin$i$-age panel of Figure~\ref{fig:Lx_age} (lower right panel). The quoted excesses are measured relative to that upper envelope boundary.

There are three obvious explanations for why some stars exhibit higher than expected \Fx/\vsini\ for their given age. First, \Fx\ may be incorrect, due to, e.g., an unresolved binary companion or flaring events that were not adequately detected and filtered out during the course of our analysis. Second, the age of the star may be incorrect, particularly if the star is older than a few Gyr and the uncertainties on the inferred stellar age are large. The third, and possibly most intriguing, possibility is that inclination effects are leading us to significantly underestimate the rotational rate of the star. We discuss each of these possibilities on a case-by-case basis for the labeled \Fx/$v$sin$i$ outliers in the lower-right panel of Figure~\ref{fig:Lx_age}.

\subsubsection{HD 192020 (G8V)}
This star exhibits an \Fx/\vsini\ excess of $\sim$55\%. There is no evidence in the literature of a spatially unresolved companion star that could be enhancing the \Lx\ attributed to HD 192020. While there is marginal evidence for temporal variability during the 31 ks \xmm\ observation of the star, there are no indications of major flaring events either in the light curve (see Figure~\ref{fig:HD192020_lc}) or in 0.3-10 keV spectrum. We therefore consider the possibility of a spatially unresolved companion or temporally unresolved flaring event biasing \Fx\ to a higher value to be unlikely. The only available age estimate of HD 192020 is 1.3$\pm$0.4 Gyr \citep{Takeda+07}, while the observed \Fx/\vsini\ value would require an age of $\sim$0.8 Gyr; it is therefore plausible that this star is somewhat younger than predicted by \citet{Takeda+07}. Finally, the observed \vsini\ of HD 192020 is low \citep[$\sim$1 km s$^{-1}$,][]{Valenti+05}. Assuming a more Sun-like rotation rate ($\sim$2 km s$^{-1}$) and an inclination angle $\sim$35\degr\ is sufficient to bring \Fx/\vsini\ into agreement with expectations, even assuming an age of 1.3 Gyr. 

\begin{figure*}
    \centering
    \includegraphics[width=1\linewidth]{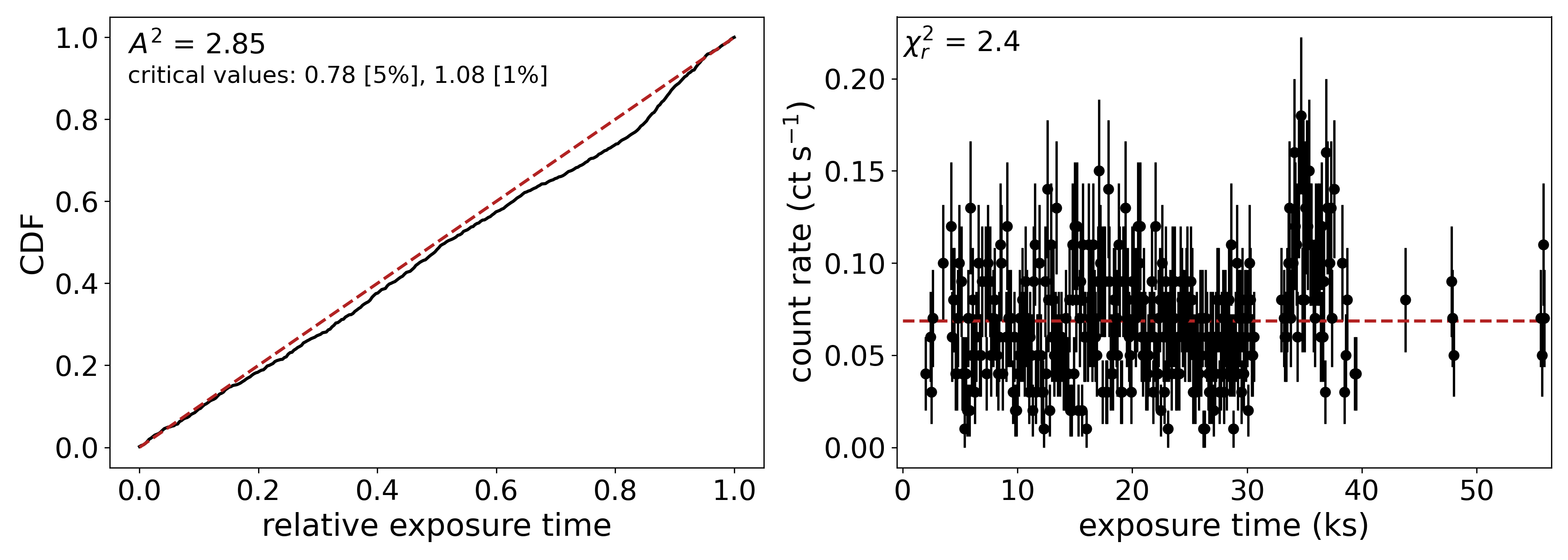}
    \caption{{\it Left}: photon arrival time cumulative distribution function of HD 192020 (\xmm\ observation ID 0721570101, black) compared to a constant count rate (red dashed line). The $A^2$ statistic and critical values are shown in the upper-left corner. {\it Right}: the light-curve data (black circles) compared to a constant median count rate (red). The reduced $\chi^2_r$ is shown in the upper-left corner. Gaps in the data at $\sim$31 ks and after $\sim40$ ks are due to background flaring events.}
    \label{fig:HD192020_lc}
\end{figure*}

\subsubsection{HD 76151 (G3V)}
The X-ray properties of HD~76151 (GL~327) were described in \citet{Binder+24}. GL~327 is an HWO Tier C target star with two available age estimates, 1.3 Gyr \citep{Takeda+07} and 2.2 Gyr \citep{Lorenzo+16,Lorenzo+18}. We adopt an intermediate age of 1.8$\pm$0.5 Gyr. This star shows an elevated \Fx/$v$sin$i$ relative to the reference envelope in Figure~\ref{fig:Lx_age}. GL~327 has a published rotation period of $P_{\rm rot}\sim15-17$ d \citep{Bellotti+25}, which corresponds to an equatorial velocity of $v_{\rm eq}\sim3.0–3.4$ km s$^{-1}$. Compared with the measured $v$sin$i=1.2$ km s$^{-1}$ \citep{Valenti+05}, this implies $i\sim20-25$\degr. Additional monitoring would still be useful to assess whether activity variability or age uncertainty also contributes to the excess.

\subsubsection{HD 219134 (K3V)}
This star, also designated GJ~892, is an HWO Tier A target star and was discussed in \citet{Binder+24}. The only age constraint available for this star is from \citet{Takeda+07}, which suggests that GJ~892 is extremely old ($>$12 Gyr). The extremely low measured \vsini\ of 0.4 km s$^{-1}$ \citep{Valenti+05} is further evidence of the old age of this star. However, the star is X-ray detected in the 0.3-10 keV band (at log\Lx$\approx$26.52, corresponding to log\Lx/\Lbol$\approx$-6.41). Although faint, the combination of this \Lx\ and \vsini\ yield an elevated \Fx/$v$sin$i$ relative to the reference envelope in Figure~\ref{fig:Lx_age}. Published rotation measurements indicate $P_{\rm rot}\approx$42 days and $i_{\star}=77\pm8$\degr\ \citep{Folsom+18}, making a low-inclination explanation unlikely. The elevated \Fx/$v$sin$i$ is therefore more likely to reflect excess activity for the adopted age, uncertainty in the age estimate, or an overestimated $>$12 Gyr age.

\subsubsection{HD 11505 (G5V)}
HD~11505 is one of the oldest stars in our sample, with three age estimates ranging from 6.2-9.3 Gyr (we adopt an age of 8.7$\pm$2.6 Gyr). It has a low \vsini\ \citep[1.5 km s$^{-1}$][]{Valenti+05}, consistent with its older age. However, it has a 0.3-10 keV log\Lx\ significantly brighter than expected for such an old star (log\Lx\ $\sim$28.2). The \Fx/\vsini\ is elevated, $\sim$118\% above the upper envelope of the gray-shaded region in Figure~\ref{fig:Lx_age}, and along with the 0.3-10 keV log\Lx\ and log\Lx/\Lbol\ ratio is consistent with a star $\sim$2 Gyr in age. Alternatively, an inclination angle of $\sim$27\degr\ could explain the observed \Fx/\vsini\ excess. Intriguingly, there is a known $\sim$9 M$_{J}$ planet in the system. \citet{Feng2022} use a combination of radial velocity and astrometry data from Gaia and Hipparcos to solve for the three dimensional architecture of the system and find an orbital inclination of $\sim$118\degr\, which is consistent with our finding (118-90=28\degr; see their Figure~A1.19). 

Another possible explanation for the high X-ray emission from HD~11505 is the presence of an unresolved, close-in binary companion. HD~11505 is listed in the Washington Double Star (WDS) catalog (WDS 01531-0120), likely due to a nearby (although likely not gravitationally bound or co-moving, on the basis of observed proper motion) M+M binary, no close stellar companion to HD~11505 has been reported in the literature. Although there was evidence of count rate variability in the archival \xmm\ observation of HD~11505, there were not a sufficient number of counts collected for spectra to be extracted and modeled. Further observations are needed to understand the nature of the elevated X-ray emission from this star.

\subsubsection{HD 130307 (K2.5V)}
There are three available age estimates for HD~130307, although these estimates range from 1.5 Gyr \citep{Takeda+07} to 2.3 Gyr \citep{Lorenzo+16,Lorenzo+18} to 8.8 Gyr \citep{Souza+24}. The absolute \Lx\ and \Lx/\Lbol\ ratios of this star are comparable to the high end of the modern solar range; we therefore adopt an age of 2.3$\pm$0.7 Gyr. HD~130307 shows a very high excess \Fx/\vsini\ of $\sim$123\%, and given the modest \vsini\ reported by \citet[][2 km s$^{-1}$]{Valenti+05} this excess could be explained by an inclination angle of $\lesssim27^{\circ}$. There is no evidence for significant X-ray count rate variability in the light curve of HD~130307.

\subsection{Relevance for HWO}
Although the majority FGK stars fall within the L$_{X}$ uncertainty envelope previously established by \citet{Gudel04}, the scatter within that envelope can exceed an order of magnitude, which can mean the difference between retention and loss of an Earth-like N$_{2}$-dominated atmosphere \citep{Tian2008,Johnstone2021b,Sherf2024}. Stars are individuals, and detailed characterization of these individual stars would best inform their likely influence on the atmospheric evolution of orbiting planets, including potential habitability and possible chemical biosignatures \citep{schwieterman+2018}. Identifying stars that show strong deviation from the canonical decay of t$^{-1.5}$ can identify age errors, which is important for properly calibrating planetary evolution models and future testing of the relationships between atmospheric composition and age via spectral characterization, such as atmospheric oxygenation \citep[e.g.][]{Bixel2020, Blunt2025, Checlair2021}. Alternatively, high \Fx/\vsini\ excess and low measured \vsini\ may indicate low projected stellar inclinations; if the planets' orbits are inclined similarly to the star, such systems may be favorable from a direct-imaging standpoint, as those planets will spend a greater time at larger angular separations from the star (and therefore outside the inner-working angle). These stars may be particularly interesting to identify when there are no known large planets detected via radial velocity measurements. 
\vspace{-0.25cm}
\section{Conclusions}\label{sec:conclusion}
We have compiled and uniformly analyzed \citep[following the methodology of][]{Binder+24} a sample of nearby Sun-like (FGK) field stars with literature age estimates and targeted \xmm\ and/or \chandra\ observations, with the goal of improving empirical constraints on coronal emission properties at ages $\gtrsim$a few Gyr where archival coverage has historically been sparse. We provide uniform 0.3-10 keV luminosities, surface fluxes, variability metrics, and (where feasible) quiescent spectral constraints for 48 nearby stars, which when combined with 37 FGK stars with age estimates from \citet{Binder+24} present a suitable sample for population-level studies of FGK coronal evolution.

We find that quiescent coronal spectra are well described by one- to three-temperature thermal plasma (\texttt{APEC}) models with characteristic plasma temperatures near $kT\sim0.1$, 0.4, and 0.8 keV, which should be interpreted as a low-order parameterization of multi-thermal coronal emission rather than as physically discrete isothermal components. Using approximate boundaries at $\sim$0.24 keV ($\sim$2.7 MK) and $\sim$0.60 keV ($\sim$7 MK), we find that the fractional contribution of the hot ($\gtrsim$7 MK) component increases strongly with X-ray surface flux: for $F_X\gtrsim10^6$ erg cm$^{-2}$ s$^{-1}$, roughly half of the 0.3-10 keV flux originates in plasma hotter than 7 MK, whereas low-$F_X$ stars are dominated by the coolest coronal plasma components.

We derive empirical power-law relations between the emission measure-weighted coronal temperature $T_{\rm EM}$ and both X-ray surface flux \Fx\ and luminosity \Lx. We find that $T_{\rm EM}\propto F_X^{0.24\pm0.05}$, consistent with prior surface flux scalings reported by \citet{Wood+18} and \citet{Johnstone+15}, and we find that $T_{\rm EM}\propto L_X^{0.27}$. These relationships provide practical temperature priors for count rate-to-flux conversions when spectroscopy is not possible, and emphasize that coronal temperature is a key variable governing observed hard-band emission. We quantify the dependence of bandpass conversions on coronal temperature, specifically between the historical soft ROSAT 0.1-2.4 keV band and the harder 0.3-10 keV band commonly adopted for \xmm. Since the 0.3-10 keV band misses a large fraction of the total \Lx, especially for cool coronae, single-temperature conversions under-predict the soft-to-hard flux ratio when the true corona is multi-thermal. We therefore utilize temperature-informed conversions from the 0.3-10 keV band to the 0.1-2.4 keV prior to interpreting apparent changes in \Lx\ with age.

Once X-ray luminosities are corrected for bandpass and temperature, the inferred 0.1-2.4 keV luminosities for our stellar sample broadly follow the canonical {\bf \Lx--age} decay relations derived from earlier work \citep[e.g., see Equation~\ref{eq:Gudel04};][]{Gudel04}, while the directly measured 0.3-10 keV luminosities exhibit substantially increased scatter at ages $\gtrsim4$ Gyr. This behavior is consistent with coronal cooling and the rapid diminution of the hottest plasma component in older, slowly rotating stars, which disproportionately affects hard-band luminosities, although larger fractional variability in low-activity stars may also contribute to the observed scatter.

We find that a small number of stars exhibit elevated \Fx/\vsini\ relative to the solar-reference age-activity envelope for their adopted ages. We discuss three plausible explanations on a case-by-case basis: unresolved companions or residual flaring biasing \Fx, large age uncertainties at late times, and/or inclination effects that lead to underestimates of the true equatorial rotation rate. Several targets (e.g., HD~11505 and HD~219134) are highlighted as high-priority systems for follow-up spectroscopy, activity monitoring, and deeper X-ray observations. Our results underscore that both the amplitude and the spectral distribution of coronal emission evolve over time. This has direct implications for reconstructing stellar irradiation histories relevant to atmospheric escape and habitability of potentially orbiting exoplanets, especially in the habitable zones of these stars, and for building consistent, mixed-mission stellar X-ray catalogs.

\begin{acknowledgments}
We would like to thank the anonymous referee, whose comments and suggestions greatly improved this manuscript. This work is supported by NASA Exoplanets Research Program (XRP) award No. 80NSSC23K0039 (PI Turnbull). This research has made use of the NASA Exoplanet Archive, which is operated by the California Institute of Technology, under contract with the National Aeronautics and Space Administration under the Exoplanet Exploration Program. This work has made use of data from the European Space Agency (ESA) mission Gaia (\url{https://www.cosmos.esa. int/gaia}), processed by the Gaia Data Processing and Analysis Consortium (DPAC, \url{https://www.cosmos.esa.int/web/gaia/dpac/consortium}). Funding for the DPAC has been provided by national institutions, in particular the institutions participating in the Gaia Multilateral Agreement. This research has made use of the SciServer science platform (\url{www.sciserver.org}). SciServer is a collaborative research environment for large-scale data-driven science. It is being developed at, and administered by, the Institute for Data Intensive Engineering and Science at Johns Hopkins University. SciServer is funded by the National Science Foundation through the Data Infrastructure Building Blocks (DIBBs) program and others, as well as by the Alfred P. Sloan Foundation and the Gordon and Betty Moore Foundation.
\end{acknowledgments}

\vspace{5mm}
\facilities{CXO, XMM, Exoplanet Archive}

\software{Astropy \citep{astropy:2013, astropy:2018, astropy:2022}}

%\bibliography{sample631}{}

\begin{thebibliography}{}
\expandafter\ifx\csname natexlab\endcsname\relax\def\natexlab#1{#1}\fi
\providecommand{\url}[1]{\href{#1}{#1}}
\providecommand{\dodoi}[1]{doi:~\href{http://doi.org/#1}{\nolinkurl{#1}}}
\providecommand{\doeprint}[1]{\href{http://ascl.net/#1}{\nolinkurl{http://ascl.net/#1}}}
\providecommand{\doarXiv}[1]{\href{https://arxiv.org/abs/#1}{\nolinkurl{https://arxiv.org/abs/#1}}}

\bibitem[{{Airapetian} {et~al.}(2017){Airapetian}, {Glocer}, {Khazanov}, {Loyd}, {France}, {Sojka}, {Danchi}, \& {Liemohn}}]{Airapetian2017}
{Airapetian}, V.~S., {Glocer}, A., {Khazanov}, G.~V., {et~al.} 2017, \apjl, 836, L3, \dodoi{10.3847/2041-8213/836/1/L3}

\bibitem[{{Arnaud}(1996)}]{Arnaud96}
{Arnaud}, K.~A. 1996, in Astronomical Society of the Pacific Conference Series, Vol. 101, Astronomical Data Analysis Software and Systems V, ed. G.~H. {Jacoby} \& J.~{Barnes}, 17

\bibitem[{{Astropy Collaboration} {et~al.}(2013){Astropy Collaboration}, {Robitaille}, {Tollerud}, {Greenfield}, {Droettboom}, {Bray}, {Aldcroft}, {Davis}, {Ginsburg}, {Price-Whelan}, {Kerzendorf}, {Conley}, {Crighton}, {Barbary}, {Muna}, {Ferguson}, {Grollier}, {Parikh}, {Nair}, {Unther}, {Deil}, {Woillez}, {Conseil}, {Kramer}, {Turner}, {Singer}, {Fox}, {Weaver}, {Zabalza}, {Edwards}, {Azalee Bostroem}, {Burke}, {Casey}, {Crawford}, {Dencheva}, {Ely}, {Jenness}, {Labrie}, {Lim}, {Pierfederici}, {Pontzen}, {Ptak}, {Refsdal}, {Servillat}, \& {Streicher}}]{astropy:2013}
{Astropy Collaboration}, {Robitaille}, T.~P., {Tollerud}, E.~J., {et~al.} 2013, \aap, 558, A33, \dodoi{10.1051/0004-6361/201322068}

\bibitem[{{Astropy Collaboration} {et~al.}(2018){Astropy Collaboration}, {Price-Whelan}, {Sip{\H{o}}cz}, {G{\"u}nther}, {Lim}, {Crawford}, {Conseil}, {Shupe}, {Craig}, {Dencheva}, {Ginsburg}, {Vand erPlas}, {Bradley}, {P{\'e}rez-Su{\'a}rez}, {de Val-Borro}, {Aldcroft}, {Cruz}, {Robitaille}, {Tollerud}, {Ardelean}, {Babej}, {Bach}, {Bachetti}, {Bakanov}, {Bamford}, {Barentsen}, {Barmby}, {Baumbach}, {Berry}, {Biscani}, {Boquien}, {Bostroem}, {Bouma}, {Brammer}, {Bray}, {Breytenbach}, {Buddelmeijer}, {Burke}, {Calderone}, {Cano Rodr{\'\i}guez}, {Cara}, {Cardoso}, {Cheedella}, {Copin}, {Corrales}, {Crichton}, {D'Avella}, {Deil}, {Depagne}, {Dietrich}, {Donath}, {Droettboom}, {Earl}, {Erben}, {Fabbro}, {Ferreira}, {Finethy}, {Fox}, {Garrison}, {Gibbons}, {Goldstein}, {Gommers}, {Greco}, {Greenfield}, {Groener}, {Grollier}, {Hagen}, {Hirst}, {Homeier}, {Horton}, {Hosseinzadeh}, {Hu}, {Hunkeler}, {Ivezi{\'c}}, {Jain}, {Jenness}, {Kanarek}, {Kendrew}, {Kern}, {Kerzendorf}, {Khvalko}, {King}, {Kirkby}, {Kulkarni},
  {Kumar}, {Lee}, {Lenz}, {Littlefair}, {Ma}, {Macleod}, {Mastropietro}, {McCully}, {Montagnac}, {Morris}, {Mueller}, {Mumford}, {Muna}, {Murphy}, {Nelson}, {Nguyen}, {Ninan}, {N{\"o}the}, {Ogaz}, {Oh}, {Parejko}, {Parley}, {Pascual}, {Patil}, {Patil}, {Plunkett}, {Prochaska}, {Rastogi}, {Reddy Janga}, {Sabater}, {Sakurikar}, {Seifert}, {Sherbert}, {Sherwood-Taylor}, {Shih}, {Sick}, {Silbiger}, {Singanamalla}, {Singer}, {Sladen}, {Sooley}, {Sornarajah}, {Streicher}, {Teuben}, {Thomas}, {Tremblay}, {Turner}, {Terr{\'o}n}, {van Kerkwijk}, {de la Vega}, {Watkins}, {Weaver}, {Whitmore}, {Woillez}, {Zabalza}, \& {Astropy Contributors}}]{astropy:2018}
{Astropy Collaboration}, {Price-Whelan}, A.~M., {Sip{\H{o}}cz}, B.~M., {et~al.} 2018, \aj, 156, 123, \dodoi{10.3847/1538-3881/aabc4f}

\bibitem[{{Astropy Collaboration} {et~al.}(2022){Astropy Collaboration}, {Price-Whelan}, {Lim}, {Earl}, {Starkman}, {Bradley}, {Shupe}, {Patil}, {Corrales}, {Brasseur}, {N{"o}the}, {Donath}, {Tollerud}, {Morris}, {Ginsburg}, {Vaher}, {Weaver}, {Tocknell}, {Jamieson}, {van Kerkwijk}, {Robitaille}, {Merry}, {Bachetti}, {G{"u}nther}, {Aldcroft}, {Alvarado-Montes}, {Archibald}, {B{'o}di}, {Bapat}, {Barentsen}, {Baz{'a}n}, {Biswas}, {Boquien}, {Burke}, {Cara}, {Cara}, {Conroy}, {Conseil}, {Craig}, {Cross}, {Cruz}, {D'Eugenio}, {Dencheva}, {Devillepoix}, {Dietrich}, {Eigenbrot}, {Erben}, {Ferreira}, {Foreman-Mackey}, {Fox}, {Freij}, {Garg}, {Geda}, {Glattly}, {Gondhalekar}, {Gordon}, {Grant}, {Greenfield}, {Groener}, {Guest}, {Gurovich}, {Handberg}, {Hart}, {Hatfield-Dodds}, {Homeier}, {Hosseinzadeh}, {Jenness}, {Jones}, {Joseph}, {Kalmbach}, {Karamehmetoglu}, {Ka{l}uszy{'n}ski}, {Kelley}, {Kern}, {Kerzendorf}, {Koch}, {Kulumani}, {Lee}, {Ly}, {Ma}, {MacBride}, {Maljaars}, {Muna}, {Murphy}, {Norman}, {O'Steen},
  {Oman}, {Pacifici}, {Pascual}, {Pascual-Granado}, {Patil}, {Perren}, {Pickering}, {Rastogi}, {Roulston}, {Ryan}, {Rykoff}, {Sabater}, {Sakurikar}, {Salgado}, {Sanghi}, {Saunders}, {Savchenko}, {Schwardt}, {Seifert-Eckert}, {Shih}, {Jain}, {Shukla}, {Sick}, {Simpson}, {Singanamalla}, {Singer}, {Singhal}, {Sinha}, {Sip{H{o}}cz}, {Spitler}, {Stansby}, {Streicher}, {{{S}}umak}, {Swinbank}, {Taranu}, {Tewary}, {Tremblay}, {Val-Borro}, {Van Kooten}, {Vasovi{'c}}, {Verma}, {de Miranda Cardoso}, {Williams}, {Wilson}, {Winkel}, {Wood-Vasey}, {Xue}, {Yoachim}, {Zhang}, {Zonca}, \& {Astropy Project Contributors}}]{astropy:2022}
{Astropy Collaboration}, {Price-Whelan}, A.~M., {Lim}, P.~L., {et~al.} 2022, \apj, 935, 167, \dodoi{10.3847/1538-4357/ac7c74}

\bibitem[{{Ayres}(2025)}]{Ayres25}
{Ayres}, T. 2025, \aj, 169, 281, \dodoi{10.3847/1538-3881/adc570}

\bibitem[{{Barnes}(2007)}]{Barnes07}
{Barnes}, S.~A. 2007, \apj, 669, 1167, \dodoi{10.1086/519295}

\bibitem[{{Bellotti} {et~al.}(2025){Bellotti}, {Petit}, {Jeffers}, {Marsden}, {Morin}, {Vidotto}, {Folsom}, {See}, \& {do Nascimento}}]{Bellotti+25}
{Bellotti}, S., {Petit}, P., {Jeffers}, S.~V., {et~al.} 2025, \aap, 693, A269, \dodoi{10.1051/0004-6361/202452378}

\bibitem[{{Binder} {et~al.}(2024){Binder}, {Peacock}, {Schwieterman}, {Turnbull}, {Virgen}, {Kane}, {Farrish}, \& {Garcia-Sage}}]{Binder+24}
{Binder}, B.~A., {Peacock}, S., {Schwieterman}, E.~W., {et~al.} 2024, \apjs, 275, 1, \dodoi{10.3847/1538-4365/ad71d6}

\bibitem[{{Bixel} \& {Apai}(2020)}]{Bixel2020}
{Bixel}, A., \& {Apai}, D. 2020, \apj, 896, 131, \dodoi{10.3847/1538-4357/ab8fad}

\bibitem[{{Blunt} {et~al.}(2025){Blunt}, {Nielsen}, {Newton}, {Christiansen}, {Daylan}, {Dressing}, {Harada}, {Kane}, {Rice}, {Mart{\'\i}nez}, \& {Sagynbayeva}}]{Blunt2025}
{Blunt}, S., {Nielsen}, E.~L., {Newton}, E.~R., {et~al.} 2025, Journal of Astronomical Telescopes, Instruments, and Systems, 11, 042214, \dodoi{10.1117/1.JATIS.11.4.042214}

\bibitem[{{Bus{\`a}} {et~al.}(2007){Bus{\`a}}, {Aznar Cuadrado}, {Terranegra}, {Andretta}, \& {Gomez}}]{Busa+07}
{Bus{\`a}}, I., {Aznar Cuadrado}, R., {Terranegra}, L., {Andretta}, V., \& {Gomez}, M.~T. 2007, \aap, 466, 1089, \dodoi{10.1051/0004-6361:20065588}

\bibitem[{{Checlair} {et~al.}(2021){Checlair}, {Villanueva}, {Hayworth}, {Olson}, {Komacek}, {Robinson}, {Popovi{\'c}}, {Yang}, \& {Abbot}}]{Checlair2021}
{Checlair}, J.~H., {Villanueva}, G.~L., {Hayworth}, B. P.~C., {et~al.} 2021, \aj, 161, 150, \dodoi{10.3847/1538-3881/abdb36}

\bibitem[{{Chmielewski}(2000)}]{Chmielewski00}
{Chmielewski}, Y. 2000, \aap, 353, 666

\bibitem[{{Claire} {et~al.}(2012){Claire}, {Sheets}, {Cohen}, {Ribas}, {Meadows}, \& {Catling}}]{Claire+12}
{Claire}, M.~W., {Sheets}, J., {Cohen}, M., {et~al.} 2012, \apj, 757, 95, \dodoi{10.1088/0004-637X/757/1/95}

\bibitem[{{Erkaev} {et~al.}(2013){Erkaev}, {Lammer}, {Odert}, {Kulikov}, {Kislyakova}, {Khodachenko}, {G{\"u}del}, {Hanslmeier}, \& {Biernat}}]{Erkaev2013}
{Erkaev}, N.~V., {Lammer}, H., {Odert}, P., {et~al.} 2013, Astrobiology, 13, 1011, \dodoi{10.1089/ast.2012.0957}

\bibitem[{{Feigelson} {et~al.}(2022){Feigelson}, {Kashyap}, \& {Siemiginowska}}]{Feigelson+22}
{Feigelson}, E.~D., {Kashyap}, V.~L., \& {Siemiginowska}, A. 2022, in Handbook of X-ray and Gamma-ray Astrophysics, 119, \dodoi{10.1007/978-981-16-4544-0_135-1}

\bibitem[{{Feng} {et~al.}(2022){Feng}, {Butler}, {Vogt}, {Clement}, {Tinney}, {Cui}, {Aizawa}, {Jones}, {Bailey}, {Burt}, {Carter}, {Crane}, {Flammini Dotti}, {Holden}, {Ma}, {Ogihara}, {Oppenheimer}, {O'Toole}, {Shectman}, {Wittenmyer}, {Wang}, {Wright}, \& {Xuan}}]{Feng2022}
{Feng}, F., {Butler}, R.~P., {Vogt}, S.~S., {et~al.} 2022, \apjs, 262, 21, \dodoi{10.3847/1538-4365/ac7e57}

\bibitem[{{Folsom} {et~al.}(2018){Folsom}, {Fossati}, {Wood}, {Sreejith}, {Cubillos}, {Vidotto}, {Alecian}, {Girish}, {Lichtenegger}, {Murthy}, {Petit}, \& {Valyavin}}]{Folsom+18}
{Folsom}, C.~P., {Fossati}, L., {Wood}, B.~E., {et~al.} 2018, \mnras, 481, 5286, \dodoi{10.1093/mnras/sty2494}

\bibitem[{{Fruscione} {et~al.}(2006){Fruscione}, {McDowell}, {Allen}, {Brickhouse}, {Burke}, {Davis}, {Durham}, {Elvis}, {Galle}, {Harris}, {Huenemoerder}, {Houck}, {Ishibashi}, {Karovska}, {Nicastro}, {Noble}, {Nowak}, {Primini}, {Siemiginowska}, {Smith}, \& {Wise}}]{Fruscione+06}
{Fruscione}, A., {McDowell}, J.~C., {Allen}, G.~E., {et~al.} 2006, in Society of Photo-Optical Instrumentation Engineers (SPIE) Conference Series, Vol. 6270, Society of Photo-Optical Instrumentation Engineers (SPIE) Conference Series, ed. D.~R. {Silva} \& R.~E. {Doxsey}, 62701V, \dodoi{10.1117/12.671760}

\bibitem[{{Fuhrmann} {et~al.}(1993){Fuhrmann}, {Axer}, \& {Gehren}}]{Fuhrmann+93}
{Fuhrmann}, K., {Axer}, M., \& {Gehren}, T. 1993, \aap, 271, 451

\bibitem[{{Gaia Collaboration} {et~al.}(2023){Gaia Collaboration}, {Vallenari}, {Brown}, {Prusti}, {de Bruijne}, {Arenou}, {Babusiaux}, {Biermann}, {Creevey}, {Ducourant}, {Evans}, {Eyer}, {Guerra}, {Hutton}, {Jordi}, {Klioner}, {Lammers}, {Lindegren}, {Luri}, {Mignard}, {Panem}, {Pourbaix}, {Randich}, {Sartoretti}, {Soubiran}, {Tanga}, {Walton}, {Bailer-Jones}, {Bastian}, {Drimmel}, {Jansen}, {Katz}, {Lattanzi}, {van Leeuwen}, {Bakker}, {Cacciari}, {Casta{\~n}eda}, {De Angeli}, {Fabricius}, {Fouesneau}, {Fr{\'e}mat}, {Galluccio}, {Guerrier}, {Heiter}, {Masana}, {Messineo}, {Mowlavi}, {Nicolas}, {Nienartowicz}, {Pailler}, {Panuzzo}, {Riclet}, {Roux}, {Seabroke}, {Sordo}, {Th{\'e}venin}, {Gracia-Abril}, {Portell}, {Teyssier}, {Altmann}, {Andrae}, {Audard}, {Bellas-Velidis}, {Benson}, {Berthier}, {Blomme}, {Burgess}, {Busonero}, {Busso}, {C{\'a}novas}, {Carry}, {Cellino}, {Cheek}, {Clementini}, {Damerdji}, {Davidson}, {de Teodoro}, {Nu{\~n}ez Campos}, {Delchambre}, {Dell'Oro}, {Esquej},
  {Fern{\'a}ndez-Hern{\'a}ndez}, {Fraile}, {Garabato}, {Garc{\'\i}a-Lario}, {Gosset}, {Haigron}, {Halbwachs}, {Hambly}, {Harrison}, {Hern{\'a}ndez}, {Hestroffer}, {Hodgkin}, {Holl}, {Jan{\ss}en}, {Jevardat de Fombelle}, {Jordan}, {Krone-Martins}, {Lanzafame}, {L{\"o}ffler}, {Marchal}, {Marrese}, {Moitinho}, {Muinonen}, {Osborne}, {Pancino}, {Pauwels}, {Recio-Blanco}, {Reyl{\'e}}, {Riello}, {Rimoldini}, {Roegiers}, {Rybizki}, {Sarro}, {Siopis}, {Smith}, {Sozzetti}, {Utrilla}, {van Leeuwen}, {Abbas}, {{\'A}brah{\'a}m}, {Abreu Aramburu}, {Aerts}, {Aguado}, {Ajaj}, {Aldea-Montero}, {Altavilla}, {{\'A}lvarez}, {Alves}, {Anders}, {Anderson}, {Anglada Varela}, {Antoja}, {Baines}, {Baker}, {Balaguer-N{\'u}{\~n}ez}, {Balbinot}, {Balog}, {Barache}, {Barbato}, {Barros}, {Barstow}, {Bartolom{\'e}}, {Bassilana}, {Bauchet}, {Becciani}, {Bellazzini}, {Berihuete}, {Bernet}, {Bertone}, {Bianchi}, {Binnenfeld}, {Blanco-Cuaresma}, {Blazere}, {Boch}, {Bombrun}, {Bossini}, {Bouquillon}, {Bragaglia}, {Bramante}, {Breedt},
  {Bressan}, {Brouillet}, {Brugaletta}, {Bucciarelli}, {Burlacu}, {Butkevich}, {Buzzi}, {Caffau}, {Cancelliere}, {Cantat-Gaudin}, {Carballo}, {Carlucci}, {Carnerero}, {Carrasco}, {Casamiquela}, {Castellani}, {Castro-Ginard}, {Chaoul}, {Charlot}, {Chemin}, {Chiaramida}, {Chiavassa}, {Chornay}, {Comoretto}, {Contursi}, {Cooper}, {Cornez}, {Cowell}, {Crifo}, {Cropper}, {Crosta}, {Crowley}, {Dafonte}, {Dapergolas}, {David}, {David}, {de Laverny}, {De Luise}, {De March}, {De Ridder}, {de Souza}, {de Torres}, {del Peloso}, {del Pozo}, {Delbo}, {Delgado}, {Delisle}, {Demouchy}, {Dharmawardena}, {Di Matteo}, {Diakite}, {Diener}, {Distefano}, {Dolding}, {Edvardsson}, {Enke}, {Fabre}, {Fabrizio}, {Faigler}, {Fedorets}, {Fernique}, {Fienga}, {Figueras}, {Fournier}, {Fouron}, {Fragkoudi}, {Gai}, {Garcia-Gutierrez}, {Garcia-Reinaldos}, {Garc{\'\i}a-Torres}, {Garofalo}, {Gavel}, {Gavras}, {Gerlach}, {Geyer}, {Giacobbe}, {Gilmore}, {Girona}, {Giuffrida}, {Gomel}, {Gomez}, {Gonz{\'a}lez-N{\'u}{\~n}ez},
  {Gonz{\'a}lez-Santamar{\'\i}a}, {Gonz{\'a}lez-Vidal}, {Granvik}, {Guillout}, {Guiraud}, {Guti{\'e}rrez-S{\'a}nchez}, {Guy}, {Hatzidimitriou}, {Hauser}, {Haywood}, {Helmer}, {Helmi}, {Sarmiento}, {Hidalgo}, {Hilger}, {H{\l}adczuk}, {Hobbs}, {Holland}, {Huckle}, {Jardine}, {Jasniewicz}, {Jean-Antoine Piccolo}, {Jim{\'e}nez-Arranz}, {Jorissen}, {Juaristi Campillo}, {Julbe}, {Karbevska}, {Kervella}, {Khanna}, {Kontizas}, {Kordopatis}, {Korn}, {K{\'o}sp{\'a}l}, {Kostrzewa-Rutkowska}, {Kruszy{\'n}ska}, {Kun}, {Laizeau}, {Lambert}, {Lanza}, {Lasne}, {Le Campion}, {Lebreton}, {Lebzelter}, {Leccia}, {Leclerc}, {Lecoeur-Taibi}, {Liao}, {Licata}, {Lindstr{\o}m}, {Lister}, {Livanou}, {Lobel}, {Lorca}, {Loup}, {Madrero Pardo}, {Magdaleno Romeo}, {Managau}, {Mann}, {Manteiga}, {Marchant}, {Marconi}, {Marcos}, {Marcos Santos}, {Mar{\'\i}n Pina}, {Marinoni}, {Marocco}, {Marshall}, {Martin Polo}, {Mart{\'\i}n-Fleitas}, {Marton}, {Mary}, {Masip}, {Massari}, {Mastrobuono-Battisti}, {Mazeh}, {McMillan}, {Messina}, {Michalik},
  {Millar}, {Mints}, {Molina}, {Molinaro}, {Moln{\'a}r}, {Monari}, {Mongui{\'o}}, {Montegriffo}, {Montero}, {Mor}, {Mora}, {Morbidelli}, {Morel}, {Morris}, {Muraveva}, {Murphy}, {Musella}, {Nagy}, {Noval}, {Oca{\~n}a}, {Ogden}, {Ordenovic}, {Osinde}, {Pagani}, {Pagano}, {Palaversa}, {Palicio}, {Pallas-Quintela}, {Panahi}, {Payne-Wardenaar}, {Pe{\~n}alosa Esteller}, {Penttil{\"a}}, {Pichon}, {Piersimoni}, {Pineau}, {Plachy}, {Plum}, {Poggio}, {Pr{\v{s}}a}, {Pulone}, {Racero}, {Ragaini}, {Rainer}, {Raiteri}, {Rambaux}, {Ramos}, {Ramos-Lerate}, {Re Fiorentin}, {Regibo}, {Richards}, {Rios Diaz}, {Ripepi}, {Riva}, {Rix}, {Rixon}, {Robichon}, {Robin}, {Robin}, {Roelens}, {Rogues}, {Rohrbasser}, {Romero-G{\'o}mez}, {Rowell}, {Royer}, {Ruz Mieres}, {Rybicki}, {Sadowski}, {S{\'a}ez N{\'u}{\~n}ez}, {Sagrist{\`a} Sell{\'e}s}, {Sahlmann}, {Salguero}, {Samaras}, {Sanchez Gimenez}, {Sanna}, {Santove{\~n}a}, {Sarasso}, {Schultheis}, {Sciacca}, {Segol}, {Segovia}, {S{\'e}gransan}, {Semeux}, {Shahaf}, {Siddiqui}, {Siebert},
  {Siltala}, {Silvelo}, {Slezak}, {Slezak}, {Smart}, {Snaith}, {Solano}, {Solitro}, {Souami}, {Souchay}, {Spagna}, {Spina}, {Spoto}, {Steele}, {Steidelm{\"u}ller}, {Stephenson}, {S{\"u}veges}, {Surdej}, {Szabados}, {Szegedi-Elek}, {Taris}, {Taylor}, {Teixeira}, {Tolomei}, {Tonello}, {Torra}, {Torra}, {Torralba Elipe}, {Trabucchi}, {Tsounis}, {Turon}, {Ulla}, {Unger}, {Vaillant}, {van Dillen}, {van Reeven}, {Vanel}, {Vecchiato}, {Viala}, {Vicente}, {Voutsinas}, {Weiler}, {Wevers}, {Wyrzykowski}, {Yoldas}, {Yvard}, {Zhao}, {Zorec}, {Zucker}, \& {Zwitter}}]{GaiaDR3}
{Gaia Collaboration}, {Vallenari}, A., {Brown}, A.~G.~A., {et~al.} 2023, \aap, 674, A1, \dodoi{10.1051/0004-6361/202243940}

\bibitem[{{Garcia-Sage} {et~al.}(2017){Garcia-Sage}, {Glocer}, {Drake}, {Gronoff}, \& {Cohen}}]{GarciaSage+17}
{Garcia-Sage}, K., {Glocer}, A., {Drake}, J.~J., {Gronoff}, G., \& {Cohen}, O. 2017, \apjl, 844, L13, \dodoi{10.3847/2041-8213/aa7eca}

\bibitem[{{Gray} {et~al.}(2006){Gray}, {Corbally}, {Garrison}, {McFadden}, {Bubar}, {McGahee}, {O'Donoghue}, \& {Knox}}]{Gray+06}
{Gray}, R.~O., {Corbally}, C.~J., {Garrison}, R.~F., {et~al.} 2006, \aj, 132, 161, \dodoi{10.1086/504637}

\bibitem[{{G{\"u}del}(1997)}]{Gudel97_flares}
{G{\"u}del}, M. 1997, \apjl, 480, L121, \dodoi{10.1086/310628}

\bibitem[{{G{\"u}del}(2004)}]{Gudel04}
---. 2004, \aapr, 12, 71, \dodoi{10.1007/s00159-004-0023-2}

\bibitem[{{G{\"u}del} {et~al.}(1997){G{\"u}del}, {Guinan}, \& {Skinner}}]{Gudel+97_age}
{G{\"u}del}, M., {Guinan}, E.~F., \& {Skinner}, S.~L. 1997, \apj, 483, 947, \dodoi{10.1086/304264}

\bibitem[{{Johnstone} {et~al.}(2021{\natexlab{a}}){Johnstone}, {Bartel}, \& {G{\"u}del}}]{Johnstone2021}
{Johnstone}, C.~P., {Bartel}, M., \& {G{\"u}del}, M. 2021{\natexlab{a}}, \aap, 649, A96, \dodoi{10.1051/0004-6361/202038407}

\bibitem[{{Johnstone} \& {G{\"u}del}(2015)}]{Johnstone+15}
{Johnstone}, C.~P., \& {G{\"u}del}, M. 2015, \aap, 578, A129, \dodoi{10.1051/0004-6361/201425283}

\bibitem[{{Johnstone} {et~al.}(2021{\natexlab{b}}){Johnstone}, {Lammer}, {Kislyakova}, {Scherf}, \& {G{\"u}del}}]{Johnstone2021b}
{Johnstone}, C.~P., {Lammer}, H., {Kislyakova}, K.~G., {Scherf}, M., \& {G{\"u}del}, M. 2021{\natexlab{b}}, Earth and Planetary Science Letters, 576, 117197, \dodoi{10.1016/j.epsl.2021.117197}

\bibitem[{Lammer {et~al.}(2007)Lammer, Lichtenegger, Kulikov, Grie{\ss}meier, Terada, Erkaev, Biernat, Khodachenko, Ribas, Penz, {et~al.}}]{Lammer2007}
Lammer, H., Lichtenegger, H.~I., Kulikov, Y.~N., {et~al.} 2007, Astrobiology, 7, 185

\bibitem[{{Linsky} {et~al.}(2020){Linsky}, {Wood}, {Youngblood}, {Brown}, {Froning}, {France}, {Buccino}, {Cranmer}, {Mauas}, {Miguel}, {Pineda}, {Rugheimer}, {Vieytes}, {Wheatley}, \& {Wilson}}]{Linsky+20}
{Linsky}, J.~L., {Wood}, B.~E., {Youngblood}, A., {et~al.} 2020, \apj, 902, 3, \dodoi{10.3847/1538-4357/abb36f}

\bibitem[{{Lorenzo-Oliveira} {et~al.}(2016){Lorenzo-Oliveira}, {Porto de Mello}, {Dutra-Ferreira}, \& {Ribas}}]{Lorenzo+16}
{Lorenzo-Oliveira}, D., {Porto de Mello}, G.~F., {Dutra-Ferreira}, L., \& {Ribas}, I. 2016, \aap, 595, A11, \dodoi{10.1051/0004-6361/201628825}

\bibitem[{{Lorenzo-Oliveira} {et~al.}(2018){Lorenzo-Oliveira}, {Freitas}, {Mel{\'e}ndez}, {Bedell}, {Ram{\'\i}rez}, {Bean}, {Asplund}, {Spina}, {Dreizler}, {Alves-Brito}, \& {Casagrande}}]{Lorenzo+18}
{Lorenzo-Oliveira}, D., {Freitas}, F.~C., {Mel{\'e}ndez}, J., {et~al.} 2018, \aap, 619, A73, \dodoi{10.1051/0004-6361/201629294}

\bibitem[{{Lyons} {et~al.}(2014){Lyons}, {Reinhard}, \& {Planavsky}}]{Lyons2014}
{Lyons}, T.~W., {Reinhard}, C.~T., \& {Planavsky}, N.~J. 2014, \nat, 506, 307, \dodoi{10.1038/nature13068}

\bibitem[{{Mamajek} \& {Hillenbrand}(2008)}]{Mamajek+08}
{Mamajek}, E.~E., \& {Hillenbrand}, L.~A. 2008, \apj, 687, 1264, \dodoi{10.1086/591785}

\bibitem[{{Matt} {et~al.}(2015){Matt}, {Brun}, {Baraffe}, {Bouvier}, \& {Chabrier}}]{Matt+15}
{Matt}, S.~P., {Brun}, A.~S., {Baraffe}, I., {Bouvier}, J., \& {Chabrier}, G. 2015, \apjl, 799, L23, \dodoi{10.1088/2041-8205/799/2/L23}

\bibitem[{Scherf \& Lammer(2021)}]{Scherf2021}
Scherf, M., \& Lammer, H. 2021, Space Science Reviews, 217, 2

\bibitem[{{Scherf} {et~al.}(2024){Scherf}, {Lammer}, \& {Spross}}]{Sherf2024}
{Scherf}, M., {Lammer}, H., \& {Spross}, L. 2024, Astrobiology, 24, e916, \dodoi{10.1089/ast.2023.0076}

\bibitem[{{Schmitt}(1997)}]{Schmitt97}
{Schmitt}, J.~H.~M.~M. 1997, \aap, 318, 215

\bibitem[{Schwieterman {et~al.}(2018)Schwieterman, Kiang, Parenteau, Harman, DasSarma, Fisher, Arney, Hartnett, Reinhard, Olson, {et~al.}}]{schwieterman+2018}
Schwieterman, E.~W., Kiang, N.~Y., Parenteau, M.~N., {et~al.} 2018, Astrobiology, 18, 663, \dodoi{10.1089/ast.2017.1729}

\bibitem[{{Shoda} {et~al.}(2020){Shoda}, {Suzuki}, {Matt}, {Cranmer}, {Vidotto}, {Strugarek}, {See}, {R{\'e}ville}, {Finley}, \& {Brun}}]{Shoda+20}
{Shoda}, M., {Suzuki}, T.~K., {Matt}, S.~P., {et~al.} 2020, \apj, 896, 123, \dodoi{10.3847/1538-4357/ab94bf}

\bibitem[{{Skumanich}(1972)}]{Skumanich72}
{Skumanich}, A. 1972, \apj, 171, 565, \dodoi{10.1086/151310}

\bibitem[{{Soderblom}(2010)}]{Soderblom10}
{Soderblom}, D.~R. 2010, \araa, 48, 581, \dodoi{10.1146/annurev-astro-081309-130806}

\bibitem[{{Souza dos Santos} {et~al.}(2024){Souza dos Santos}, {Porto de Mello}, {Costa-Bhering}, {Lorenzo-Oliveira}, {Almeida-Fernandes}, {Dutra-Ferreira}, \& {Ribas}}]{Souza+24}
{Souza dos Santos}, P.~V., {Porto de Mello}, G.~F., {Costa-Bhering}, E., {et~al.} 2024, \mnras, 532, 563, \dodoi{10.1093/mnras/stae1532}

\bibitem[{{Stassun} {et~al.}(2019){Stassun}, {Oelkers}, {Paegert}, {Torres}, {Pepper}, {De Lee}, {Collins}, {Latham}, {Muirhead}, {Chittidi}, {Rojas-Ayala}, {Fleming}, {Rose}, {Tenenbaum}, {Ting}, {Kane}, {Barclay}, {Bean}, {Brassuer}, {Charbonneau}, {Ge}, {Lissauer}, {Mann}, {McLean}, {Mullally}, {Narita}, {Plavchan}, {Ricker}, {Sasselov}, {Seager}, {Sharma}, {Shiao}, {Sozzetti}, {Stello}, {Vanderspek}, {Wallace}, \& {Winn}}]{TESSinput}
{Stassun}, K.~G., {Oelkers}, R.~J., {Paegert}, M., {et~al.} 2019, \aj, 158, 138, \dodoi{10.3847/1538-3881/ab3467}

\bibitem[{{Stelzer} {et~al.}(2007){Stelzer}, {Flaccomio}, {Briggs}, {Micela}, {Scelsi}, {Audard}, {Pillitteri}, \& {G{\"u}del}}]{Stelzer+07}
{Stelzer}, B., {Flaccomio}, E., {Briggs}, K., {et~al.} 2007, \aap, 468, 463, \dodoi{10.1051/0004-6361:20066043}

\bibitem[{{Takeda} {et~al.}(2007){Takeda}, {Ford}, {Sills}, {Rasio}, {Fischer}, \& {Valenti}}]{Takeda+07}
{Takeda}, G., {Ford}, E.~B., {Sills}, A., {et~al.} 2007, \apjs, 168, 297, \dodoi{10.1086/509763}

\bibitem[{{Telleschi} {et~al.}(2005){Telleschi}, {G{\"u}del}, {Briggs}, {Audard}, {Ness}, \& {Skinner}}]{Telleschi+05}
{Telleschi}, A., {G{\"u}del}, M., {Briggs}, K., {et~al.} 2005, \apj, 622, 653, \dodoi{10.1086/428109}

\bibitem[{{Testa} {et~al.}(2015){Testa}, {Saar}, \& {Drake}}]{Testa+15}
{Testa}, P., {Saar}, S.~H., \& {Drake}, J.~J. 2015, Philosophical Transactions of the Royal Society of London Series A, 373, 20140259, \dodoi{10.1098/rsta.2014.0259}

\bibitem[{{Tian} {et~al.}(2008){Tian}, {Kasting}, {Liu}, \& {Roble}}]{Tian2008}
{Tian}, F., {Kasting}, J.~F., {Liu}, H.-L., \& {Roble}, R.~G. 2008, Journal of Geophysical Research (Planets), 113, E05008, \dodoi{10.1029/2007JE002946}

\bibitem[{{Tu} {et~al.}(2015){Tu}, {Johnstone}, {G{\"u}del}, \& {Lammer}}]{Tu+15}
{Tu}, L., {Johnstone}, C.~P., {G{\"u}del}, M., \& {Lammer}, H. 2015, \aap, 577, L3, \dodoi{10.1051/0004-6361/201526146}

\bibitem[{{Turnbull}(2015)}]{Turnbull15}
{Turnbull}, M.~C. 2015, arXiv, \dodoi{10.48550/arXiv.1510.01731}

\bibitem[{{Vaiana} {et~al.}(1981){Vaiana}, {Cassinelli}, {Fabbiano}, {Giacconi}, {Golub}, {Gorenstein}, {Haisch}, {Harnden}, {Johnson}, {Linsky}, {Maxson}, {Mewe}, {Rosner}, {Seward}, {Topka}, \& {Zwaan}}]{Vaiana+81}
{Vaiana}, G.~S., {Cassinelli}, J.~P., {Fabbiano}, G., {et~al.} 1981, \apj, 245, 163, \dodoi{10.1086/158797}

\bibitem[{{Valenti} \& {Fischer}(2005)}]{Valenti+05}
{Valenti}, J.~A., \& {Fischer}, D.~A. 2005, \apjs, 159, 141, \dodoi{10.1086/430500}

\bibitem[{{Voges}(1993)}]{Voges93}
{Voges}, W. 1993, Advances in Space Research, 13, 391, \dodoi{10.1016/0273-1177(93)90147-4}

\bibitem[{{Voges} {et~al.}(1999){Voges}, {Aschenbach}, {Boller}, {Br{\"a}uninger}, {Briel}, {Burkert}, {Dennerl}, {Englhauser}, {Gruber}, {Haberl}, {Hartner}, {Hasinger}, {K{\"u}rster}, {Pfeffermann}, {Pietsch}, {Predehl}, {Rosso}, {Schmitt}, {Tr{\"u}mper}, \& {Zimmermann}}]{Voges+99}
{Voges}, W., {Aschenbach}, B., {Boller}, T., {et~al.} 1999, \aap, 349, 389, \dodoi{10.48550/arXiv.astro-ph/9909315}

\bibitem[{{Wood} {et~al.}(2018){Wood}, {Laming}, {Warren}, \& {Poppenhaeger}}]{Wood+18}
{Wood}, B.~E., {Laming}, J.~M., {Warren}, H.~P., \& {Poppenhaeger}, K. 2018, \apj, 862, 66, \dodoi{10.3847/1538-4357/aaccf6}

\bibitem[{{Wright} {et~al.}(2011){Wright}, {Drake}, {Mamajek}, \& {Henry}}]{Wright+11}
{Wright}, N.~J., {Drake}, J.~J., {Mamajek}, E.~E., \& {Henry}, G.~W. 2011, \apj, 743, 48, \dodoi{10.1088/0004-637X/743/1/48}

\end{thebibliography}
%\bibliographystyle{aasjournal}

\end{document}